\documentclass[pra,onecolumn,floatfix,superscriptaddress,longbibliography,notitlepage, nofootinbib]{revtex4-1}
\pdfoutput=1

\usepackage[utf8]{inputenc}
\usepackage{braket}
\usepackage{amsmath}
\DeclareMathOperator{\Tr}{Tr}
\usepackage{dcolumn}   
\usepackage{bm}        
\usepackage{amssymb}
\usepackage{mathtools}
\usepackage{tikz}
\usepackage{qcircuit}
\usepackage{appendix}
\usepackage{hyperref}
\usepackage{subcaption}
\usepackage{amsmath,amsfonts,amssymb}
\usepackage{mathtools}
\usepackage{thmtools,thm-restate}
\usepackage{algorithm}
\usepackage{mathrsfs}
\usepackage{tikz}
\usepackage{subcaption}
\usepackage{pgfplots}
\usetikzlibrary{3d}
\usepackage{tkz-euclide}

\usepackage{algpseudocode}

\newcommand{\xbf}{\textbf{x}}
\newcommand{\Hbf}{\textbf{H}}
\newcommand{\Ibb}{\mathbb{I}}
\newcommand{\Hsr}{\mathscr{H}}
\newcommand{\Nsr}{\mathscr{N}}

\newtheorem{definition}{Definition}
\newtheorem{theorem}{Theorem}
\newtheorem{lemma}{Lemma}

\usetikzlibrary{arrows.meta}

\def\BibTeX{{\rm B\kern-.05em{\sc i\kern-.025em b}\kern-.08em
    T\kern-.1667em\lower.7ex\hbox{E}\kern-.125emX}}

\begin{document}
\title{Quantum Algorithm For Testing Convexity of Function}
\author{Nhat A. Nghiem}
\email{nhatanh.nghiemvu@stonybrook.edu}
\affiliation{Department of Physics and Astronomy, State University of New York at Stony Brook, Stony Brook, NY 11794-3800, USA}
\affiliation{C. N. Yang Institute for Theoretical Physics, State University of New York at Stony Brook, Stony Brook, NY 11794-3840, USA}

\author{Tzu-Chieh Wei}
\affiliation{Department of Physics and Astronomy, State University of New York at Stony Brook, Stony Brook, NY 11794-3800, USA}
\affiliation{C. N. Yang Institute for Theoretical Physics, State University of New York at Stony Brook, Stony Brook, NY 11794-3840, USA}
\begin{abstract}
Functions are a fundamental object in mathematics, with countless applications to different fields, and are usually classified based on certain properties, given their domains and images. An important property of a real-valued function is its convexity, which plays a very crucial role in many areas, such as thermodynamics and geometry. Motivated by recent advances in quantum computation as well as the quest for quantum advantage, we give a quantum algorithm for testing convexity of polynomial functions, which appears frequently in multiple contexts, such as optimization, machine learning, physics, etc. We show that quantum computers can reveal the convexity property superpolynomially faster than classical computers with respect to number of variables. As a corollary, we provide a significant improvement and extension on quantum Newton's method constructed in earlier work of Rebentrost et al [New J. Phys. \textbf{21} 073023 (2019)]. We further discuss our algorithm in a broader context, such as potential application in the study of geometric structure of manifold, testing training landscape of variational quantum algorithm and also gradient descent/Newton's method for optimization. 
\end{abstract}
\maketitle

\section{Introduction}
Quantum computers hold great promise to solve difficult computational problems that lie beyond the reach of classical computers. The underlying power of quantum computers is due to two intrinsic properties of quantum mechanics: superposition and entanglement. Tremendous efforts have been made to exploit the potential of quantum computers in various contexts. Some early pioneering works~\cite{deutsch1985quantum, deutsch1992rapid} showed that quantum computers could probe properties of a blackbox function with a single query usage. The breakthrough work of Shor~\cite{shor1994proceedings} showed that quantum computers can factorize a given integer number superpolynomially faster than their classical counterpart, which has been recently improved in~\cite{regev2023efficient}. Grover~\cite{grover1996fast} later showed that a quadratic speedup is achieved for unstructured database search. Further quantum speedup has been showcased in a wide array of problems, such as simulating quantum systems~\cite{berry2007efficient, berry2012black, berry2014high, berry2015hamiltonian,low2017optimal, low2019hamiltonian}, solving linear systems~\cite{harrow2009quantum, childs2017quantum}, supervised and unsupervised learning ~\cite{lloyd2013quantum, mitarai2018quantum}, principle component analysis~\cite{lloyd2014quantum}, topological data analysis~\cite{lloyd2016quantum}, learning from experiments~\cite{huang2022quantum}, etc. As a whole, these developments have ignited an exciting view towards the application of quantum computers, as well as triggering efforts in experimental realization of fault-tolerance devices as to bring quantum computation steps closer to reality. \\  

Given these successes, the question of the repertoire of tasks quantum computers can excel in is still worthy of pursuing. An important model that has been central to investigating quantum advantage is the so-called \textit{blackbox model}. In such a model, we have access to the blackbox with an unknown structure, that computes some Boolean functions, e.g., accepting some input strings and then outputting a Boolean variable (0 or 1). As the structure is unknown to us, the goal is to extract properties of such functions with minimum resource, e.g., the number of access to the corresponding blackbox. In fact, some early works~\cite{deutsch1985quantum,deutsch1992rapid} showed that quantum computers can reveal hidden properties of given functions using a minimum number of queries much smaller compared to classical counterparts. More relevant to our work, Jordan~\cite{jordan2005fast} considered the numerical gradient estimation problem given the blackbox that computed some function explicitly and showed that a single query is sufficient to reveal the gradient at a given point, up to some desired accuracy. Inspired by such a line of pursuit, we consider the potential advantage of quantum computers in the topic of functional analysis, where our object of interest is a (multivariate) function, which is a very basic object in mathematics. The intriguing point is that a function can possess very rich analytical properties, and thus the problem is very appealing to explore from a computational perspective. A particular property that we focus on in our work is the \textit{convexity}, which captures the shape of the function in some domain (see Fig.~\ref{fig:mainfigure} and~\ref{fig: twovari} below).
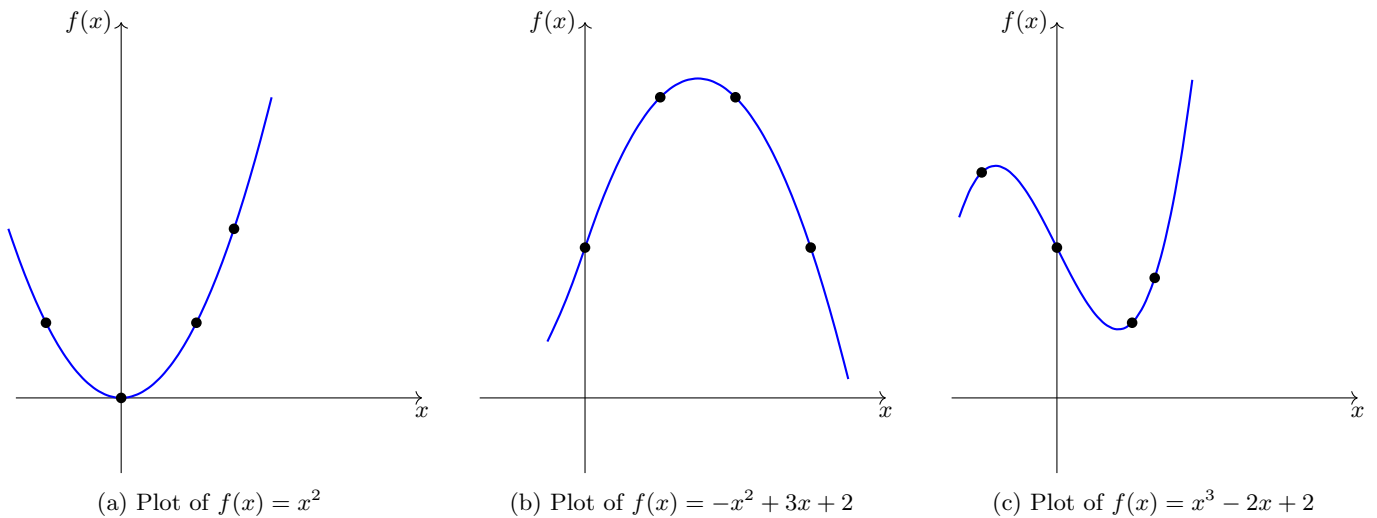
\begin{figure}[h]
    \centering
    \begin{subfigure}[b]{0.3\textwidth}
    \centering
        \begin{tikzpicture}
            
\draw[->] (-1.4,0) -- (4,0) node[below] {$x$};
\draw[->] (0,-1) -- (0,5) node[left] {$f(x)$};

\draw[blue, thick, domain=-1.5:2, smooth, variable=\x] plot (\x, {\x*\x});

\foreach \i in {-1,0,1,1.5}
    \fill (\i,{\i*\i}) circle (2pt) node[above right] {};
        \end{tikzpicture}
        \caption{Plot of $f(x) = x^2$}
        \label{fig:subfig1}
    \end{subfigure}
    \hfill
    \begin{subfigure}[b]{0.3\textwidth}
    \centering
        \begin{tikzpicture}
            
\draw[->] (-1.4,0) -- (4,0) node[below] {$x$};
\draw[->] (0,-1) -- (0,5) node[left] {$f(x)$};

\draw[blue, thick, domain=-0.5:3.5, smooth, variable=\x] plot (\x, {-\x^2 + 3*\x + 2});

\foreach \i in {0,1,2,3}
    \fill (\i,{-\i^2 + 3*\i + 2}) circle (2pt) node[above right] {};
        \end{tikzpicture}
        \caption{Plot of $f(x) = -x^2+3x+2$}
        \label{fig:subfig2}
    \end{subfigure}
    \hfill
    \begin{subfigure}[b]{0.3\textwidth}
    \centering
        \begin{tikzpicture}
            
\draw[->] (-1.4,0) -- (4,0) node[below] {$x$};
\draw[->] (0,-1) -- (0,5) node[left] {$f(x)$};
\draw[blue, thick, domain=-1.3:1.8, smooth, variable=\x] plot (\x, {\x^3 - 2*\x + 2});

\foreach \i in {-1,0,1,1.3}
    \fill (\i,{\i^3 - 2*\i + 2}) circle (2pt)  node[above right] { };
        \end{tikzpicture}
        \caption{Plot of $f(x) = x^3 -2x+2$}
        \label{fig:subfig3}
    \end{subfigure}
    \caption{Some illustrations for convexity (1 variable case). The function $f(x) = x^2$ (Fig. \ref{fig:subfig1}) is convex for all $\mathbb{R}$. The function $f(x) = -x^2+3x+2$ (Fig \ref{fig:subfig2}) is strictly not convex, or concave for whole domain $\mathbb{R}$. The function $f(x)=x^3-2x+2$ (see Fig.~\ref{fig:subfig3}) is only convex in some domain.  }
    \label{fig:mainfigure}
\end{figure}
A convex function has a peculiar feature that a local minimum is also a global minimum (see Fig.~\ref{fig:subfig1} for simple illustration). Additionally, if a function is convex in some domain, then a minimum is easily obtained, e.g., by the gradient descent method, which makes it very useful in optimization areas.  

In this work, we aim to tackle the challenge of testing the convexity of some polynomial function. We begin with a simple case, which is a homogeneous polynomial of even degree (to be defined later) and subsequently, building upon such homogeneous polynomial, we generalize the construction to arbitrary polynomial type. The kind of homogeneous polynomial of even degree was also considered in~\cite{rebentrost2019quantum}, where the authors proposed the quantum gradient descent and quantum Newton's method for finding local minima. In fact, our work is quite inspired by them, as the structure of a given function makes it simpler to compute two important quantities: the gradient and Hessian. Building upon~\cite{rebentrost2019quantum}, we show that the ability to obtain the analytical form of Hessian translates into the ability to test convexity by examining the sign of Hessian's spectrum, and that quantum computers can achieve that goal with superpolynomially less cost than their classical counterpart in relative to the number of variables included in the given polynomial function. Along the way, as a corollary, we show how to construct the Hessian more efficiently than the original method in~\cite{rebentrost2019quantum}, thus providing a significant improvement on quantum Newton's method that also appeared in \cite{rebentrost2019quantum}. First, our improved quantum Newton's method work on arbitrary polynomial, instead of homogeneous one of even degree. Second, our method reduces the complexity dependence on (inverse of) error tolerance from polynomial to polylogarithmic. The complexity dependence on the degree of given function is also reduced by a power of 4. Additionally, at each step of the Newton's method, the number of copies of (block encoding of) temporal solution required in our work is polynomially less than (by a power of 5) that of \cite{rebentrost2014quantum}. We further mention three potential subjects that might be useful for our framework: differential geometry, variational quantum algorithm and gradient descent as well as Newton's method for finding minima of objective function.  \\

The structure of the paper is as follows. First, in Section~\ref{sec: overview}, we provide an overview of our objectives, with the corresponding assumptions and criteria for convexity in Section~\ref{sec: dissecting}. Our main algorithm is outlined in details in Section~\ref{sec: qalgorithm}. Remarks and further discussions are given in Section~\ref{sec: remarkanddiscussion}, where we show that our method can be generalized to an arbitrary polynomial and showcase some potential applications of our method in the context of differential geometry and variational quantum algorithm, as well as improving the quantum Newton's method originally introduced in~\cite{rebentrost2019quantum}. Appendix ~\ref{sec: prelim} contains the necessary recipes that underlie our work. Appendix~\ref{sec: alphadiagonal} provides a proof of Lemma~\ref{lemma: alphadiagonal}.

\section{Overview}
\subsection{Overview of the Problem and Assumptions}
\label{sec: overview}
We consider a multivariate function $f: \mathbb{R}^n \longrightarrow \mathbb{R}$ which is a homogeneous polynomial of an even degree $2p$ ($p \in \mathbb{Z}$). Let $\textbf{x} = (x_1, x_2, ..., x_n)$; as shown in~\cite{rebentrost2018quantum}, such a polynomial admits a tensor algebraic decomposition:
\begin{align}
    f(\textbf{x}) = \frac{1}{2} \textbf{x}^T \otimes \cdots \otimes \textbf{x}^T A \,\textbf{x} \otimes \cdots \otimes \textbf{x},
\end{align}
where $A$ is an $s$-sparse matrix of dimension $n^p \times n^p$. 
\begin{figure}[h]
    \centering
    \begin{tikzpicture}
    \begin{axis}[
        colormap={whiteblue}{color(0cm)=(white); color(1cm)=(blue)},
        view={60}{30},     
        xmin=-2, xmax=2,
        ymin=-2, ymax=2,
        zmin=0, zmax=6,
        xtick={-2,-1,...,2},
        ytick={-2,-1,...,2},
        ztick={0,1,...,6},
        grid=both,
        minor tick num=1,
        axis lines=middle,
        axis line style={-latex},
        width=10cm,
        height=8cm,
        enlarge y limits=true,
        enlarge x limits=true,
        enlarge z limits=true,
    ]
    \addplot3[
        surf,
        shader=interp,
        domain=-2:2,
        domain y=-2:2,
    ] {x^2 + x*y + y^2};
    \end{axis}
\end{tikzpicture}
    \caption{Example of a homogeneous polynomial of degree 2: $f(x,y) = x^2 + xy + y^2$.}
    \label{fig: twovari}
\end{figure}
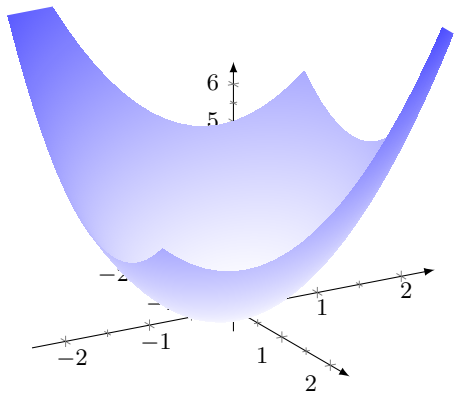
Throughout this work, we assume that the oracle's access to $A$ is given in a similar fashion to that in~\cite{rebentrost2019quantum}. The knowledge of the polynomial is obtained by the oracle's access to matrix $A$, which can be formally decomposed as:
\begin{align}
    A = \sum_{\alpha = 1}^K A_1^\alpha \otimes \cdots \otimes A_p^\alpha,
\end{align}
where each $A_i^\alpha$ is a matrix of size $n\times n$ ($i= 1,2,...,p$) and $K$ is some natural number. While the above general decomposition is not required explicitly in our work (i.e., we do not assume the oracle has direct access to the submatrices $A_i$'s), the expression is helpful as the important quantities, such as the gradient and Hessian, admit analytical forms. More specifically, the gradient can be written as:
\begin{align}
\label{eqn: gradient}
    \Vec{\nabla} f(\textbf{x}) = D(\textbf{x}) \xbf,
\end{align}
where $D$ is specified as 
\begin{align}
    D(\xbf) =  \sum_{\alpha=1}^K\sum_{j=1}^p \Big( \prod_{i=1, i\neq j}^p  \xbf^T A_i^{\alpha} \xbf \Big) A_j^{\alpha},
\end{align}
and the Hessian matrix of function $f$ can be written as: 
\begin{align}
\label{eqn: hessian}
    \Hbf (\xbf) = 2 \sum_{\alpha =1}^K \,\sum_{j,k = 1, j\neq k}^p \,\,\prod_{i=1, i\neq j,k}^p (\xbf^T A_i^\alpha \xbf)   A_k^\alpha \xbf \xbf^T A^\alpha_j + D(\xbf).
\end{align}
 The special formulation above provides us an alternative way to compute the Hessian and gradient at a given point $\xbf$ as follows. Denote $\xbf \xbf^T$ as $\rho_x$, then we can write the gradient as
\begin{align}
    D(\xbf) = \Tr_{1,2,...,p-1} \big( M_D \ ( (\xbf\xbf^T)^{\otimes p-1} \otimes \Ibb_n  )  \big), 
    \label{eqn: 6}
\end{align}
where 
\begin{align}
    M_D = \sum^p_{m=1} M_m,
\end{align}
and each $M_m$ ($m=1,..,p)$ can be obtained from $A$ via permutation of entries as explained below. Recall that 
\begin{align}
    A = \sum_{\alpha = 1}^K A_1^\alpha \otimes \cdots\otimes A_m^\alpha\otimes \cdots \otimes A_p^\alpha,
\end{align}
and $M_m$ is defined similarly to $A$ except having the $p$-th matrix swapped with the $m$-th matrix, i.e.,
\begin{align}
    M_m = \sum_{\alpha = 1}^K A_1^\alpha \otimes \cdots  \otimes A_p^\alpha \otimes \cdots \otimes  A_m^\alpha.
\end{align}
In a similar fashion, the Hessian can be expressed alternatively as:
\begin{align}
    \Hbf(\xbf) = \Tr_{1,2,...,p-1} \big( (M_H + M_D)  \ ( (\xbf\xbf^T) ^{\otimes p-1} \otimes \Ibb_n )   \big),
    \label{eqn: 10}
\end{align}
where 
\begin{align}
    M_H = 2 \sum_{j \neq k}^p \Theta_{jk},
\end{align}
and each $\Theta_{jk}$ (for $j,k = 1,...,p)$ can be obtained from $A$ via permutation of matrices, e.g., the $j,k$-th matrices are swapped to $p-1$ and $p$-th ones, respectively, 
\begin{align}
    \Theta_{jk} = \sum_{\alpha = 1}^K A_1^\alpha \otimes \cdots \otimes A_{p-1}^\alpha \otimes \cdots \otimes A_p^\alpha \otimes \cdots\otimes A_j^\alpha \otimes A_k^\alpha.
\end{align}
Therefore, the oracle access to $A$ allows us to obtain entries of all $\Theta_{jk}$ respectively. \\

We remark that in the above, we discuss homogeneous polynomials of even degree as a primary target due to its simplification of gradient and Hessian formulation. As mentioned in~\cite{rebentrost2019quantum}, inhomogeneity can be inserted. For example, given a homogeneous polynomial $f_{\rm homo}$, one can multiply it with $c^T \xbf $ where $c$ is some $n$-dimensional vector to obtain an inhomogeneous polynomial. In the following, we describe our framework with a homogeneous polynomial of even degree and then provide a generalization in Section~\ref{sec: generalization}. 

Let $||\cdot||$ denotes the operator norm. As specified in~\cite{rebentrost2019quantum}, we have that the norm $|| D || \leq p ||A||$ and $||\Hbf || \leq p^2 ||A||$. Without loss of generalization, we can set the norm of $A$, $||A|| = 1$, so that $p^2 ||A|| \leq p^2$, and thus we can guarantee that $||D|| < p$ and $||\Hbf|| \leq p^2$. Throughout this work, we assume this condition holds for convenience, as rescaling by some factor does not change the nature of the problem, e.g., the convexity of function and particularly the (asymptotical) running time of our method. 

\subsection{Dissecting Convexity of Function}
\label{sec: dissecting}

We are interested in the convexity of $f$ in some domain $\mathscr{D} \subset \mathbb{R}^n$. By trivially redefining the function (e.g., with a coordinate shift), we can choose $\mathscr{D}$ to be a hypersphere with radius $1$ for simplicity. The following criterion is a well-known result in mathematics and can be found in many standard literature. \\

\noindent
\textbf{Criterion:} \textit{Over some domain $\mathscr{D}$, if the Hessian matrix of a given function $f$ is positive-semidefinite at every point, then the function $f$ is convex.} \\

The application of the above criterion is straightforward. We compute the Hessian matrix of $f$ at chosen points in the region and check its spectrum. The positive-semidefiniteness of a matrix is equivalent to all eigenvalues being non-negative. Therefore, the sign of the spectrum is necessary and sufficient for revealing convexity. As we mentioned, the Hessian matrix $\Hbf$ can be written down explicitly, which is convenient for spectral analysis. As the dimension of $\Hbf$ grows with the number of variables (there are $n$ variables), finding the full spectrum of $\Hbf$ is computationally demanding. Furthermore, the only information that matters is the sign of the smallest eigenvalue, as we only need to see if it is positive or not. Therefore, we propose to find the \textit{minimum} eigenvalue of $\Hbf$. If such an eigenvalue is non-negative, then all other eigenvalues are non-negative, which implies that $\Hbf$ is positive-semidefinite and hence, the corresponding $f$ is convex within the domain $\mathscr{D}$. In the following section, we construct a quantum algorithm to first find and then verify the sign of the smallest eigenvalue, thereby dissecting the convexity of the function $f$, given the details of $f$ and related assumptions from the previous section Sec.~\ref{sec: overview}. 

\section{Quantum Algorithm }
\label{sec: qalgorithm}
We begin with a remark that all details regarding relevant definitions as well as useful tools are provided in Appendix~\ref{sec: prelim}. Here proceed to describe our main result.

\subsection{Constructing Hessian $\Hbf$ for a Single Point}
\label{sec: singlepointhessian}
In Sec.~\ref{sec: dissecting}, we first introduce the general idea behind our work: finding the minimum eigenvalue of $\Hbf$. Our first task is to produce the block encoding of $\Hbf$ that can be used for further analysis. In order to achieve this goal, we recall the following important formulation:
\begin{align}
    \Hbf(\xbf) = \Tr_{1,2,...,p-1} \big( (M_H+M_D)  \ ( (\xbf\xbf^T)^{\otimes p-1} \otimes \Ibb_n )   \big),
\end{align}
where $ M_H = 2 \sum_{j \neq k}^p \Theta_{jk}$ 
 
and $M_D = \sum_{m=1}^p M_m$. The oracle access to entries of each $M_m$ and $H_{jk}$ allows us to use Lemma~\ref{lemma: As} to produce the $\epsilon$-approximated block encoding of $M_H/(sp^2)$ and $M_D/(sp)$ (where $s$ is the sparsity of $A$) in time: 
$  \mathcal{O}\Big(  p\big(p\log(n) + \log^{2.5}(\frac{1}{\epsilon})\big )  \Big)$. 
Thus, it is quite simple to obtain the block encoding of $M_D/(sp^2)$ from $M_D/(sp)$  by multiplying with a factor $1/p$ (see Lemma~\ref{lemma: scale}). Lemma~\ref{lemma: sumencoding} then allows us to obtain the $\epsilon$-approximated block encoding of $(M_H+M_D)/(2sp^2)$. 

The next recipe that we need includes the following simple relations.
\begin{align}
  &  \Tr_1 (A \otimes B) (\xbf \xbf^T \otimes \Ibb) = (\xbf^T A \xbf) B, \\
 &   (\xbf\xbf^T \otimes \Ibb) (A \otimes B) (\xbf\xbf^T \otimes \Ibb) = \xbf\xbf^T \otimes (\xbf^T A \xbf) B.
\end{align}
Using these properties plus the tensor structure of $M_D$ and $M_H$, we can show that:
\begin{align}
\label{eqn: singlehessian}
    (\xbf\xbf^T) ^{\otimes p-1} \otimes \Ibb_n \ (M_H + M_D) \ (\xbf\xbf^T) ^{\otimes p-1} \otimes \Ibb_n &= (\xbf\xbf^T) ^{\otimes p-1} \otimes \Hbf(\xbf).
\end{align}
The reason comes from the fact that
\begin{align}
    (\xbf\xbf^T)^{\otimes p-1} = \xbf^{\otimes p-1} (\xbf^T)^{\otimes p-1}
\end{align}
and that
\begin{align}
    (\xbf^T)^{\otimes p-1} \otimes \Ibb_n \ (M_H+M_D) \ \xbf^{\otimes p-1} \otimes \Ibb_n = \Hbf(\xbf).
\end{align}
For now, we assume that we have a block encoding of $(\xbf\xbf^T)$, as subsequently, we will show how to produce such a block encoding and generalize it to deal with multiple points chosen from the domain $\mathscr{D}$, as required by the convexity criterion. Lemma~\ref{lemma: tensorproduct} allows us to use $p-1$ block encodings of $(\xbf\xbf^T)$ and a trivial block encoding of $\Ibb_n$ to obtain the block encoding of $(\xbf\xbf^T)^{\otimes p-1} \otimes \Ibb_n$. Then Lemma \ref{lemma: product} (combined with the simple relations derived above) yields the block encoding of $(\xbf\xbf^T)^{\otimes p-1} \otimes \Hbf(\xbf)/(2sp^2)$. 

\subsection{Constructing ``multi-points'' Hessian}
\label{sec: multipointhessian}
\subsubsection{For homogeneous polynomial of even degree}
We remark that the above construction yields the block encoding of the operator $(\xbf\xbf^T)^{\otimes(p-1)} \otimes \Hbf (\xbf)/(2sp^2)$, which contains the tensor product of $\xbf \xbf^T$ and Hessian matrix $\Hbf$ at a given point $\xbf$. As it will become clearer later, in order to take advantage of quantum parallelism for analyzing the Hessian spectrum at multiple points, we need to adjust the above procedure to construct what we call a ``multi-point'' Hessian. \\

For further clarity and to avoid confusion with the above construction, we set the following notation for subsequent discussions. Let $\mathscr{N}$ be the number of points of consideration; $\xbf_i$ ($\in \mathbb{R}^n$) is the $i$-th point; the corresponding Hessian of $f$ evaluated at $\xbf_i$ is then $\Hbf(\xbf_i)$. The first goal is to produce the block encoding of the following operator $ \bigoplus_i^{\mathscr{N}} (\xbf_i\xbf_i^T)^{ \otimes {p-1} } \otimes \Hbf(\xbf_i)/(2sp^2) $, which has the matrix representation as:
\begin{align}
  \frac{1}{2sp^2} \  \begin{pmatrix}
         (\xbf_1\xbf_1^T)^{ \otimes {p-1} } \otimes \Hbf(\xbf_1) & \cdots  & \cdots & \cdots \\
         \cdots &  (\xbf_2\xbf_2^T)^{ \otimes {p-1} } \otimes \Hbf(\xbf_2) & \cdots & \cdots \\
         \cdots & \cdots &  \cdots  & \cdots  \\
         \cdots & \cdots & \cdots &  (\xbf_\mathscr{N} \xbf_{\mathscr{N}}^T)^{ \otimes {p-1} } \otimes \Hbf(\xbf_\mathscr{N} )  \\
    \end{pmatrix}.
\end{align}
Now, we outline the procedure that produces the desired block encoding. First, we have the following lemma:
\begin{lemma}
\label{lemma: multimatrix}
   Given block encoding of $(M_H+M_D)/(2sp^2)$ (as constructed in Sec.~\ref{sec: singlepointhessian}), the block encoding of operator $\bigoplus_i^{\mathscr{N}} (M_H + M_D)/(2sp^2)$ can be prepared with extra $\mathcal{O}(1)$ cost.
\end{lemma}
\textit{Proof:} To show this, we need to add an extra register of dimension $\mathscr{N}$. The resulting tensor product of $\Ibb_{\mathscr{N}}$ and unitary block encoding of $(M_H+M_D)/(2sp^2)$ produces the block encoding of $\Ibb_{\mathscr{N}} \otimes (M_H+M_D)/(2sp^2)$, which is exactly $\bigoplus_i^{\mathscr{N}} (M_H + M_D)/(2sp^2)$ by a simple algebraic property. \\

Next, we introduce the following crucial recipe:
\begin{lemma}
\label{lemma: multipointstate}
    The block encoding of the operator $ \bigoplus_{i=1}^{\mathscr{N}} (\xbf_i\xbf_i^T)^{\otimes p-1} \otimes \Ibb_n$ can be prepared in time $\mathcal{O}(p(\log(n\Nsr))$.
\end{lemma}
\textit{Proof:} To prove the above lemma, we first consider a $\log(n\mathscr{N})$ qubits system which resides a Hilbert space $\mathscr{H}$ of dimension $n\mathscr{N}$. Let $U$ be some unitary operator. Once acting on the basis state $\ket{\bf 0}_\Hsr$ we obtain the state $\ket{\phi}$, i.e.,
\begin{align}
\label{eqn: U}
    U  \ket{\bf 0}_\Hsr = \ket{\phi}.
\end{align}
Let $\ket{\phi} = (x_1,x_2,...,x_{n\mathscr{N}})/C$ where
$$ C = \sqrt{ \sum_{k=1}^{n\Nsr} x_k^2 } $$
is the normalization factor. In this paper, we work in the real regime, i.e., $x_i \in \mathbb{R}$ for all $i = 1,2,..., n\Nsr$.  If we break such a vector into $\Nsr$ parts, and denote each part as $\xbf_j$ ($j=1,2,..., \Nsr)$, e.g., $\xbf_j = (x_{(j-1)n +1},x_{(j-1)n+2}, ..., x_{(j-1)n +n})$. Note that given such notation, the normalization factor $C$ is essentially equal to $C = \sum_{i=1}^\Nsr |\xbf_i|^2$ where $|.|^2$ refers to the usual $l_2$ norm of a vector. It is then straightforward to decompose $\ket{\phi}$ as: $\ket{\phi} = \frac{1}{C} \sum_{j=1}^{ \mathscr{N}} \ket{j} \otimes \xbf_j $. Before proceeding, we remark on the unitary $U$ that generates the desired state $\ket{\phi}$. We are interested in $\Nsr$ points $\{\xbf_i\}_{i=1}^\Nsr$. If we choose these $\Nsr$ points classically, which means that we know their coordinates plus the norms respectively, then we can use the well-known amplitude encoding method~\cite{schuld2018supervised, prakash2014quantum} to load these entries into a quantum state, resulting in a quantum circuit $U$ of depth $\mathcal{O}(\log(n\Nsr))$. On the other hand, if we choose $U$ to be a random unitary circuit, then a constraint is imposed on the coordinates of $\ket{\phi}$, which implies $C=1$. In this case, the set of points $\{\xbf_i\}_{i=1}^\Nsr$ must have their norms summing up to 1, which may reduce the number of points considered in a given domain in one go. 

Now we append another ancillary system having dimension $\Nsr$ initialized in $\ket{\bf 0}_{\Nsr}$, to obtain the state $\ket{\bf 0}_{\Nsr} \otimes \ket{\phi} = \frac{1}{C} \sum_{j=1}^{ \mathscr{N}}\ket{\bf 0}_{\Nsr} \otimes \ket{j} \otimes \xbf_j $. Using CNOT gates to copy the second register to the first one, i.e.,
\begin{align}
  \frac{1}{C}\sum_{j=1}^{ \mathscr{N}}\ket{\bf 0}_{\Nsr}  \otimes \ket{j} \otimes \xbf_j \longrightarrow  \frac{1}{C}\sum_{j=1}^{ \mathscr{N}}\ket{j}_{\Nsr} \otimes \ket{j} \otimes \xbf_j. 
\end{align}
If we trace out the first register (with subscript $\Nsr$), we obtain the state $(1/C^2)\sum_{j=1}^\Nsr \ket{j}\bra{j} \ \otimes \ \xbf_j\xbf_j^T$ (we recall that it should be $\xbf_j\xbf_j^\dagger$, but since we work in the real regime, $\xbf^T$ and $\xbf^\dagger$ are identical). Lemma~\ref{lemma: improveddme} allows us to prepare the (exact) block encoding of $(1/C^2) \sum_{j=1}^\Nsr \ket{j}\bra{j} \ \otimes \ \xbf_j\xbf_j^T$ (in complexity $\mathcal{O}(\log (n\Nsr) )$, whose matrix representation is as follows:
\begin{align}
   \frac{1}{C^2} \sum_{j=1}^\Nsr \ket{j}\bra{j} \ \otimes \ \xbf_j\xbf_j^T = \frac{1}{C^2}\begin{pmatrix}
        \xbf_1\xbf_1^T & \cdot & \cdot &\cdot \\
        \cdot & \xbf_2 \xbf_2^T & \cdot & \cdot \\
        \cdot & \cdot & \cdot & \cdot & \\
        \cdot & \cdot & \cdot & \xbf_\Nsr \xbf_\Nsr^T 
    \end{pmatrix}.
    \label{eqn: C}
\end{align}
If $C$ is greater than $1$, the factor $C^2$ from the above representation can be removed using the amplification technique (basically uniform singular value amplification from~\cite{gilyen2019quantum}), with a further complexity $\mathcal{O}(C^2)$. In reality, if we prefer to choose $\Nsr$ points $\{\xbf_i\}_{i=1}^\Nsr$ uniformly within the hypersphere, then we can expect that $C^2 = \sum_{i=1}^\Nsr |\xbf_i|^2 \leq \Nsr$, which means that the complexity of the above step can be $\mathcal{O}(\Nsr)$ (for $C \geq 1$). 

For $C$ smaller than $1$, one cannot use the amplification method~\cite{gilyen2019quantum}.  Therefore, the factor $C$ cannot be removed by amplification. (Of course, this can be avoided using different or more points.) For now, we continue the construction with $C$ being greater or equal to $1$ (which means it is removed from the above equation). Subsequently, we will return to the case $C$ being smaller than $1$ and show that the structure of a homogeneous polynomial allows us to factor out some power of $C$, resulting in a different expression for the Hessian evaluated at given points. Hence, the final complexity is different (from the case $C> 1$) by a factor of power of $C$. 

Define $\ket{\xbf_i} \equiv \xbf_i/|\xbf_i|$, e.g., the normalized vector. We have that $\xbf_i\xbf_i^T = |\xbf_i|^2 \ket{\xbf_i}\bra{\xbf_i}$. Our goal is to transform the above operator into its square root, e.g., for each $i$, we aim to transform $|\xbf_i|^2 \ket{\xbf_i}\bra{\xbf_i} \longrightarrow |\xbf_i|\ket{\xbf_i}\bra{\xbf_i}$. In order to achieve such a goal, we recall two results from~\cite{gilyen2019quantum} and~\cite{gilyen2019quantum1}:
\begin{lemma}[Corollary 3.4.14 in \cite{gilyen2019quantum1}]
\label{lemma: positivepower}
    Given $\delta, \epsilon \in (0, 1/2]$, $c\in (0,1]$ and let $f(x) = 0.5 x^c$. There exists an even/odd polynomial of degree $\mathcal{O}( \frac{1}{\delta} \log( \frac{1}{\epsilon} ) )$ such that
    $$  ||P-f||_{ [\delta,1] } \leq \epsilon, \ ||P||_{[-1,1]} \leq 1. $$
\end{lemma}

\begin{lemma}\label{lemma: qsvt}[\cite{gilyen2019quantum} Theorem 56]
\label{lemma: theorem56}
Suppose that $U$ is an
$(\alpha, a, \epsilon)$-encoding of a Hermitian matrix $A$. (See Definition 43 of \cite{gilyen2019quantum} for the definition).
If $P \in \mathbb{R}[x]$ is a degree-$d$ polynomial satisfying that
\begin{itemize}
\item for all $x \in [-1,1]$: $|P(x)| \leq \frac{1}{2}$,
\end{itemize}
then, there is a quantum circuit $\tilde{U}$, which is an $(1,a+2,4d \sqrt{\frac{\epsilon}{\alpha}})$-encoding of $P(A/\alpha)$ and
consists of $d$ applications of $U$ and $U^\dagger$ gates, a single application of controlled-$U$, and $\mathcal{O}((a+1)d)$
other one- and two-qubit gates.
\end{lemma}
Choosing $c= 1/2$, by using polynomial from Lemma~\ref{lemma: positivepower} as an approximation to $0.5\sqrt{x}$ plus Lemma~\ref{lemma: theorem56}, we can obtain the following ($\epsilon$-approximated) transformation (remind that for all $i$, $\xbf_i\xbf_i^T \equiv |\xbf_i|^2 \ket{\xbf_i}\bra{\xbf_i}$):
\begin{align}
    \begin{pmatrix}
        \xbf_1\xbf_1^T & \cdot & \cdot &\cdot \\
        \cdot & \xbf_2 \xbf_2^T & \cdot & \cdot \\
        \cdot & \cdot & \cdot & \cdot & \\
        \cdot & \cdot & \cdot & \xbf_\Nsr \xbf_\Nsr^T 
    \end{pmatrix} \longrightarrow  \begin{pmatrix}
        0.5 |\xbf_1|  \ket{\xbf_1}\bra{\xbf_1} & \cdot & \cdot &\cdot \\
        \cdot & 0.5 |\xbf_2| \ket{\xbf_2}\bra{\xbf_2} & \cdot & \cdot \\
        \cdot & \cdot & \cdot & \cdot & \\
        \cdot & \cdot & \cdot & 0.5 |\xbf_\Nsr| \ket{\xbf_\Nsr} \bra{\xbf_\Nsr} 
    \end{pmatrix}
    \label{eqn: squared}
\end{align}
Note that we can not remove the factor 0.5 by amplification, as the norm needs to be less than or equal to 1/2. Moreover, Lemma~\ref{lemma: positivepower} admits an approximation of a given positive power function on the interval $[\delta,1]$. To apply such a result to our context, we need to make sure that our interval is appropriate, which means that $\delta$ needs to be less than or equal to $\min_i \{ |\xbf_i|^2 \}_{i=1}^\Nsr$. In order to find such a minimum, we can use Lemma~\ref{lemma: findingmin} to the left-hand side of the above equation in the case that the points are unknown. However, if we classically pick these points, we know their norms, including the minimum.

In Sec.~\ref{sec: prelim}, we have mentioned a very simple way to prepare the block encoding of the identity matrix of any dimension. Suppose we have a block encoding of $\Ibb_{n(p-1)}$, then Lemma~\ref{lemma: tensorproduct} allows us to prepare the block encoding of $\sum_{i=1}^\Nsr \ket{i}\bra{i} \otimes |\xbf_i|\ket{\xbf_i}\bra{\xbf_i} \otimes \Ibb_n^{\otimes p-1}$. We note that with $2\log(n)$ SWAP gates, we can swap the order of $|\xbf_i|\ket{\xbf_i}\bra{\xbf_i}$ and any $\Ibb_n$ among $p-1$ such ones. Therefore, it takes $p-1$ further steps to achieve all of them, e.g., block encoding of operators of the form 
$$ \sum_{i=1}^\Nsr \ket{i}\bra{i} \otimes \Ibb_n \otimes |\xbf_i| \ket{\xbf_i}\bra{\xbf_i} \otimes \cdots \otimes \Ibb_n, \ \sum_{i=1}^\Nsr \ket{i}\bra{i} \otimes \Ibb_n \otimes \Ibb_n \otimes |\xbf_i|\ket{\xbf_i}\bra{\xbf_i} \otimes \cdots \otimes \Ibb_n, \ ..., \sum_{i=1}^\Nsr \ket{i}\bra{i} \otimes \Ibb_n \otimes \Ibb_n \otimes \cdots \otimes |\xbf_i|\ket{\xbf_i}\bra{\xbf_i} \otimes \Ibb_n.$$

Lemma~\ref{lemma: product} yields the block encoding of their products, and it is easy to note that their product is $\sum_{i=1}^\Nsr \ket{i}\bra{i} \otimes ( |\xbf_i|\ket{\xbf_i}\bra{\xbf_i})^{\otimes p-1} \otimes \Ibb_n$, which is exactly $ \bigoplus_{i=1}^{\mathscr{N}} (|\xbf_i|\ket{\xbf_i}\bra{\xbf_i})^{\otimes p-1} \otimes \Ibb_n $ by a simple tensor algebraic property. The complexity of this step is $\mathcal{O}(p\log(n\Nsr) \frac{1}{|\xbf_{min}|})$, where $|\xbf_{min}| = \min_i \{ |\xbf_i|^2 \}_{i=1}^\Nsr$. To summarize what we have so far, we state the following:
\begin{lemma}
 Assuming that $C = \sqrt{ \sum_{i=1}^\Nsr |\xbf_i|^2} \geq 1$. An $\epsilon$-approximated block encoding of $\bigoplus_{i=1}^{\mathscr{N}} (|\xbf_i|\ket{\xbf_i}\bra{\xbf_i})^{\otimes p-1} \otimes \Ibb_n$ can be prepared in time complexity $\mathcal{O}( p \frac{1}{|\xbf_{min}|}\log(n\Nsr) \log(\frac{1}{\epsilon})  ).$ 
\end{lemma}

We are now ready to construct the ``multi-point'' Hessian, which is straightforward by combining the methods outlined in~\ref{sec: singlepointhessian}, Lemmas~\ref{lemma: multimatrix} and~\ref{lemma: multipointstate} and the following simple property of matrix multiplication:
\begin{align}
    \begin{pmatrix}
        A_1 & 0 \\
        0 & A_2 
    \end{pmatrix} 
    \cdot 
    \begin{pmatrix}
        B_1 & 0 \\
        0 & B_2 
    \end{pmatrix} 
    =  \begin{pmatrix}
        A_1B_1 & 0 \\
        0 & A_2B_2 
    \end{pmatrix},
\end{align}
which holds for any higher dimension $N$, i.e.,
\begin{align}
\label{eqn: tensorproperty}
    \bigoplus_{i=1}^N A_i \ \bigoplus_{i=1}^N B_i = \bigoplus_{i=1}^N A_i B_i.
\end{align}
More specifically, from Lemma~\ref{lemma: multimatrix}, we have  the block encoding of $\bigoplus_{i=1}^\Nsr (M_H+M_D)/(2sp^2)$. Using Lemma \ref{lemma: product} to obtain the block encoding of 
$$ \bigoplus_{i=1}^{\mathscr{N}} \Big( 0.5|\xbf_i|\ket{\xbf_i}\bra{\xbf_i} \Big) ^{\otimes p-1} \otimes \Ibb_n  \ \cdot \ \bigoplus_{i=1}^\Nsr \frac{(M_H+M_D)}{2sp^2} \cdot  \ \bigoplus_{i=1}^{\mathscr{N}} \Big( 0.5 |\xbf_i| \ket{\xbf_i}\bra{\xbf_i} \Big)^{\otimes p-1} \otimes \Ibb_n, $$
which is exactly 
\begin{align*}
    \bigoplus_{i=1}^\Nsr  \Big( 0.5|\xbf_i|\ket{\xbf_i}\bra{\xbf_i} \Big)^{\otimes p-1} \otimes \Ibb_n  \cdot \frac{(M_H+M_D)}{ 2sp^2} \cdot \Big( 0.5|\xbf_i|\ket{\xbf_i}\bra{\xbf_i} \Big)^{\otimes p-1} \otimes \Ibb_n &= \frac{1}{4^{p-1}} \bigoplus_{i=1}^\Nsr (\ket{\xbf_i}\bra{\xbf_i})^{\otimes p-1} \otimes \frac{\Hbf(\xbf_i)}{2sp^2}.
    \label{eqn: bigsumHessian}
\end{align*}
Due to the fact that, for all $i$, $|\xbf_i|\ket{\xbf_i}\bra{\xbf_i} = \ket{\xbf_i} \xbf_i^T$ (basically we absorb the norm to the $\bra{\xbf_i}$ to obtain $\xbf_i^T$), then we use property~(\ref{eqn: singlehessian}) to obtain the block encoding of $\bigoplus_{i=1}^\Nsr (\ket{\xbf_i}\bra{\xbf_i})^{\otimes p-1} \otimes \Hbf(\xbf_i)/(2sp^2)$. We remark a subtlety that, for any $i$, $0.5|\xbf_i|\ket{\xbf_i}$ is essentially proportional to $|\xbf_i|\ket{\xbf_i}$, which means that one can `absorb' that factor $0.5$ into the calculation of Hessian. This is a particular property of homogeneous polynomial (see Sec.~\ref{sec: overview}), whereby a rescale of the given input $\xbf \longrightarrow \lambda \xbf$ (for $\lambda \in \mathbb{R}$) would result in a rescaling of Hessian, i.e., $\Hbf(\lambda \xbf) = \lambda^{2(p-1)} \Hbf(\xbf)$. 

At the beginning of this section, we mentioned two cases: the normalization factor $C \geq 1$ or $C < 1$. For $C <1$, we cannot remove the factor $C$ in Eqn~(\ref{eqn: C}); the aforementioned property of homogeneous polynomial then allows us to handle the case $C \leq 1$ in a simple manner, as we treat the point $\xbf_i/C$ as a scaled point $\xbf_i \longrightarrow \xbf_i/C$, which means that eventually there will be a factor $C^{2p-2}$ absorbed, e.g, in the above formulation, we would have the following operator: 
$$ \frac{1}{(4C^2)^{p-1}} \bigoplus_{i=1}^\Nsr (\ket{\xbf_i}\bra{\xbf_i})^{\otimes p-1} \otimes \frac{\Hbf(\xbf_i)}{2sp^2}. $$
    
One might wonder that for the case $C \geq 1$ if we did not use amplification to remove the factor $C$, then we would end up having the same form as above. It does not seem to be an issue for homogeneous polynomials, as discussed above, due to the difference in the Hessians being just a scaling factor. However, there are two reasons. First, our method is aimed at dealing with polynomials of arbitrary type, as we will generalize subsequently. This means that the homogeneous property will not hold; the homogeneous polynomial is just a base on which we can build and achieve the generalization conveniently. Second, as mentioned previously, for $C < 1 $, we can not use the amplification method to remove the factor $C$. Besides that, the subsequent strategy that we will use to dissect convexity (in Section~\ref{sec: dissecting}) is tracking the value of some operators that have the form as the above. If there is an extra factor $C$ (being greater than 1), we will need to choose the error tolerance to be smaller (by a factor of $C^{2p-2}$) to reveal the correct eigenvalue of the desired Hessian, which would result in a substantial running time.

With the above operator (for both cases $C \geq 1$ and $C < 1$), we are able to generalize it to polynomials of arbitrary type. The following section shows how to achieve our goal based on what we have obtained so far.

\subsubsection{Generalization to Polynomial of Arbitrary Kind}
\label{sec: generalization}
Previously, we used the particular form of $f$, which is a homogeneous polynomial of even degree. Now we generalize our method to deal with arbitrary polynomials, or more specifically, monomials as also mentioned in~\cite{rebentrost2019quantum}, which include homogeneous polynomials of odd degree and inhomogeneous polynomials. According to~\cite{rebentrost2019quantum}, an inhomogeneous function can be given below by inserting an extra factor into a homogeneous one:
\begin{align}
    f(\xbf) = \sum_{q=1}^{P-1} (c_q^T \xbf) \  \prod_{k=1}^{q-1} (\xbf^T B_{k q } \xbf).
\end{align}
We recognize that the term $  \prod_{k=1}^{q-1} (\xbf^T B_{kq} \xbf)$ is in fact an alternative expression for a homogeneous polynomial of even degree $2(q-1)$, and $c_q^T \xbf$ is a $\xbf$-dependent coefficient that adds inhomogeneities. Therefore, all terms in the above summation share a similar form, and the function $f(x)$ is simply adding them. Since the derivative of a sum of functions is the sum of the derivative of the constituting functions, we consider the following part separately $g(\xbf)  =  (c^T \xbf) \  \prod (\xbf^T B \xbf)$ where we already ignore the subscript and treat with full generalization, e.g., with arbitrary order of the polynomial. The partial derivative:
\begin{align}
    \frac{\partial g}{\partial x_m} = c^T \xbf \frac{\partial \prod (\xbf^T B \xbf)  } { \partial x_m } + \prod (\xbf^T B \xbf)  \frac{\partial c^T \xbf}{ \partial x_m}.
\end{align}
Now we take further partial derivative:
\begin{align}
    \frac{\partial^2 g}{\partial x_n \partial x_m} = c^T\xbf \ \frac{\partial^2 \prod (\xbf^T B \xbf)  } {\partial x_n \partial x_m } + \frac{\partial \prod (\xbf^T B \xbf)  } { \partial x_m } \frac{\partial c^T \xbf}{\partial x_n} + \frac{\partial \prod (\xbf^T B \xbf)  } { \partial x_n }  \frac{\partial c^T \xbf}{ \partial x_m}.
\end{align}
Denote $\prod (\xbf^T B \xbf)  \equiv h(\xbf)$. Since the term $\prod (\xbf^T B \xbf) $ is a homogeneous polynomial of even degree, we know its gradient (composed of partial derivatives, see Eqn.~(\ref{eqn: gradient})) and Hessian matrix (composed of partial derivatives of second order, see Eqn.~(\ref{eqn: hessian})), therefore, the above equation implies that the Hessian of $g$ is generally expressed as:
\begin{align}
    \Hbf(g(\xbf)) = (c^T \xbf) \  \Hbf(h(\xbf) ) +  \Vec{\nabla} h(\xbf)  c^T + c  \Vec{\nabla} h(\xbf) ^T.
    \label{eqn: Hinhomo}
\end{align}
Since $h(\xbf)$ is a homogeneous polynomial of even degree, its gradient and Hessian can be computed using known technique~\cite{rebentrost2019quantum} (see further Sec.~\ref{sec: overview}, more specifically Eqn~(\ref{eqn: gradient}) and Eqn~(\ref{eqn: hessian})). The only difference we need to consider is the contribution from $c$, which accounts for the inhomogeneities part. How to deal with the extra term $c^T \xbf$ depends greatly on what kind of access we have to $c$. We now outline a solution in the case where $c$ is generated by some unitary $U_C$, e.g., $U_C \ket{\bf 0} = \ket{c} \equiv c$ (assuming further that $|c|=1$). From the unitary $U_C$, using Lemma~\ref{lemma: improveddme} we have the block encoding of $cc^T$.  \\

For now, we assume to work in the regime where the normalization factor $C \geq 1$. The first goal is to produce the block encoding of some operator that includes $\Vec{\nabla} h(\xbf) c^T$ as a tensor component (as in Sec.~\ref{sec: multipointhessian}). Since $h(\xbf)$ is the regular homogeneous part, the gradient operator can be computed according to Sec.~\ref{sec: overview}, i.e., there exists a procedure similar to what was outlined in Sec.~\ref{sec: singlepointhessian} and Sec.~\ref{sec: multipointhessian} that produces the block encoding of the following operator  
$$ \bigoplus_{i=1}^\Nsr \frac{1}{4^{p-1}}(\ket{\xbf_i}\bra{\xbf_i})^{\otimes p-1} \otimes \frac{1}{sp^2} \Vec{\nabla} h(\xbf_i) \xbf_i^T. $$
More specifically, from equation~(\ref{eqn: singlehessian}), if we ignore the $M_H$ operator, we then obtain a similar property for $D(\xbf)$:
\begin{align}
    (\xbf\xbf^T) ^{\otimes p-1} \otimes \Ibb_n \cdot M_D \cdot (\xbf\xbf^T) ^{\otimes p-1} \otimes \Ibb_n &= (\xbf\xbf^T) ^{\otimes p-1} \otimes D(\xbf).
\end{align}
Consequently, we can use the result of Section \ref{sec: singlepointhessian} and \ref{sec: multipointhessian} to obtain the block encoding of the operator $\bigoplus_{i=1}^\Nsr \Big( 0.5 |\xbf_i|\ket{\xbf_i}\bra{\xbf_i}\Big)^{p-1}\otimes \Ibb_n$. Then we use the block encoding of operator $M_D/sp^2$ plus lemma \ref{lemma: product} to obtain the block encoding of their products, which is basically 
$$ \frac{1}{4^{p-1}} \bigoplus_{i=1}^\Nsr  (\ket{\xbf_i}\bra{\xbf_i})^{\otimes p-1} \otimes \frac{1}{sp^2} D(\xbf_i).  $$
From Sec.~\ref{sec: multipointhessian}, we have the operator $\bigoplus_{i=1}^\Nsr \xbf_i\xbf_i^T$ (the construction above Lemma 3). If we use Lemma~\ref{lemma: tensorproduct} to construct the block encoding of $\bigoplus_{i=1}^\Nsr \xbf_i \xbf_i^T \otimes \Ibb_n^{\otimes p-1}$. Then, with $2\log(n)$ SWAP gates, we can swap the first and last register, e.g., obtaining $\bigoplus_{i=1}^\Nsr \Ibb_n^{\otimes p-1} \otimes \xbf_i\xbf_i^T$. Then we use Lemma~\ref{lemma: product} to obtain the block encoding of:
\begin{align}
    \frac{1}{4^{p-1}} \bigoplus_{i=1}^\Nsr  (\ket{\xbf_i}\bra{\xbf_i})^{\otimes p-1} \otimes \frac{1}{sp^2} D(\xbf_i)   \cdot \bigoplus_{i=1}^\Nsr \Ibb_n^{\otimes p-1} \otimes \xbf_i\xbf_i^T &= \frac{1}{4^{p-1}} \bigoplus_{i=1}^\Nsr (\ket{\xbf_i}\bra{\xbf_i})^{\otimes p-1} \otimes \frac{1}{sp^2}  D(\xbf_i)\xbf_i\xbf_i^T \\
    &=  \bigoplus_{i=1}^\Nsr \frac{1}{4^{p-1}}(\ket{\xbf_i}\bra{\xbf_i})^{\otimes p-1} \otimes \frac{1}{sp^2} \Vec{\nabla} h(\xbf_i) \xbf_i^T.
\end{align}
Note that we used the property of gradient operator of homogeneous even degree function $D(\xbf_i)\xbf_i = \Vec{\nabla} h(\xbf_i)$. \\

From the block encoding of $cc^T$, it is trivial to produce the block encoding of $\bigoplus_{i=1}^\Nsr \Ibb_{n}^{\otimes p-1} \otimes cc^T$. Denote $c^T \xbf_i = \xbf_i^T c \equiv \beta_i$. We then use Lemma~\ref{lemma: product} to produce the block encoding of the multiplied operator:
\begin{align}
    \bigoplus_{i=1}^\Nsr ( \ket{\xbf_i}\bra{\xbf_i})^{\otimes p-1} \otimes \frac{\Vec{\nabla} h(\xbf_i) \xbf_i^T}{4^{p-1} sp^2} \cdot  \bigoplus_{i=1}^\Nsr \Ibb_{n}^{\otimes p-1} \otimes cc^T &= \bigoplus_{i=1}^\Nsr (\ket{\xbf_i}\bra{\xbf_i})^{\otimes p-1} \otimes \frac{\beta_i}{4^{p-1} sp^2} \Vec{\nabla} h(\xbf_i) c^T.
\end{align}
In order to remove the factor $\beta_i$ from the above formulation, we use the following procedure. 

In Appendix~\ref{sec: alphadiagonal}, we prove the following thing:
\begin{lemma}
   \label{lemma: alphadiagonal}
   Given the block encoding of $cc^T$, then it is possible to construct the block encoding of the diagonal matrix $\mathscr{B}$ with entries $\mathscr{B}_{ij} = \beta_i\beta_j \delta_{ij}$ where $\beta_i = \xbf_i^T c $.
\end{lemma}
Recall that for the homogeneous part $h(\xbf)$, procedure in Sec.~\ref{sec: multipointhessian} obtains the following operator:
$$  \frac{1}{4^{p-1}} \bigoplus_{i=1}^\Nsr (\ket{\xbf_i}\bra{\xbf_i})^{\otimes p-1} \otimes \frac{1}{2sp^2} \Hbf(\xbf_i). $$
Thus, we use Lemma~\ref{lemma: alphadiagonal} and Lemma~\ref{lemma: product} to obtain the block encoding of:
$$ \frac{1}{4^{p-1}} \bigoplus_{i=1}^\Nsr \ket{\xbf_i}\bra{\xbf_i}^{\otimes p-1 } \otimes \frac{ \beta_i^2 \Hbf(\xbf_i) }{2sp^2}.$$
Recall that we have the block encoding of the operator 
$$  \bigoplus_{i=1}^\Nsr (\ket{\xbf_i}\bra{\xbf_i})^{\otimes p-1} \otimes \frac{\beta_i}{4^{p-1} sp^2} \Vec{\nabla} h(\xbf_i) c^T.$$
Using Lemma~\ref{lemma: scale} to insert an extra factor $1/2$ into the above operator, i.e., we obtain: 
$$ \bigoplus_{i=1}^\Nsr (\ket{\xbf_i}\bra{\xbf_i})^{\otimes p-1} \otimes \frac{\beta_i}{4^{p-1} 2 sp^2} \Vec{\nabla} h(\xbf_i) c^T.$$
We remark that, the transpose of the block encoding of the above operator is the block encoding of: 
$$ \bigoplus_{i=1}^\Nsr (\ket{\xbf_i}\bra{\xbf_i})^{\otimes p-1} \otimes \frac{\beta_i}{4^{p-1} 2 sp^2} c \Vec{\nabla}  h(\xbf_i)^T  $$

We then use Lemma~\ref{lemma: sumencoding} to obtain the block encoding of a sum of two operators above, 
\begin{align}
    \mathbb{P} &= \frac{1}{3} \Big(  \frac{1}{4^{p-1}} \bigoplus_{i=1}^\Nsr \ket{\xbf_i}\bra{\xbf_i}^{\otimes p-1 } \otimes \frac{ \beta_i^2 \Hbf(\xbf_i) }{2sp^2} +  \bigoplus_{i=1}^\Nsr (\ket{\xbf_i}\bra{\xbf_i})^{\otimes p-1} \otimes \frac{\beta_i}{4^{p-1} 2 sp^2} \Vec{\nabla} h(\xbf_i) c^T +  \\
    & \bigoplus_{i=1}^\Nsr (\ket{\xbf_i}\bra{\xbf_i})^{\otimes p-1} \otimes \frac{\beta_i}{4^{p-1} 2 sp^2} c \Vec{\nabla}  h(\xbf_i)^T\Big) \\
    &= \frac{1}{3} \frac{1}{4^{p-1}sp^2} \Big(  \bigoplus_{i=1}^\Nsr \beta_i \ket{\xbf_i}\bra{\xbf_i}^{\otimes p-1 } \otimes \Hbf_{\text{inho}}(\xbf_i)    \Big),
\end{align}
where $\Hbf_{\text{inho}}(\xbf_i) $ refers generally to the Hessian evaluated at $\xbf_i$ of the given inhomogeneous function (see derivation in Eqn. \ref{eqn: Hinhomo}). Our goal is to remove the factor $\beta_i$ (for all $i$) in the above operator. Recall that from Lemma~\ref{lemma: alphadiagonal} we have block encoding of $\mathscr{B}$ that contains $\beta_i^2$ (for $i=1,2,...,\Nsr$) on the diagonal. Note that we also have block encoding of $\mathscr{B} \otimes \Ibb_n  \equiv \bigoplus_{i=1}^\Nsr \beta_i^2 \Ibb_n $ (which is trivial to obtain using block encoding of identity matrix $\Ibb_n$ plus Lemma~\ref{lemma: tensorproduct}). We then can use the following polynomial approximation of the negative power function from~\cite{gilyen2019quantum1}:
\begin{lemma}[Corollary 3.4.13 in~\cite{gilyen2019quantum1}]
    Let $\delta, \epsilon \in (0,1/2], c>0$ and let $f(x) = \frac{\delta^c}{2} x^{-c}$, then there exists a (could be even or odd) polynomial $P$ such that $|| P - f(x) ||_{\delta,1} \leq \epsilon, ||P||_{-1,1} \leq 1/2$. The degree of the polynomial $P$ is $\mathcal{O}(\frac{\max [1,c]}{\delta} \log(\frac{1}{\epsilon})  )$.
\end{lemma}
To apply the above lemma to the operator $\bigoplus_{i=1}^\Nsr \beta_i^2 \Ibb_n  $, we need to know the lower bound, i.e., the minimum of $\{\beta_i^2\}_{i=1}^\Nsr$, denoted as $\beta_{min}$ which can be estimated using Lemma~\ref{lemma: findingmin}. Then, we  use the above Lemma (with $c = 1/2$) with Lemma~\ref{lemma: qsvt} to $\epsilon$ approximately transform the operator 
$$ \bigoplus_{i=1}^\Nsr \beta_i^2 \Ibb_n \longrightarrow  \bigoplus_{i=1}^\Nsr \frac{\sqrt{\beta_{min}}}{2\beta_i} \Ibb_n.  $$
The complexity of this step is $\mathcal{O}( \frac{1}{\beta_{min}} \log(\frac{1}{\epsilon} ) )$. Then, we can use Lemma~\ref{lemma: product} to obtain the approximation of the block encoding of 
$$ \bigoplus_{i=1}^\Nsr \frac{ \sqrt{\beta_{min}} }{2\beta_i} \Ibb_n \cdot  \mathbb{P}  =  \frac{\sqrt{\beta_{min}}}{2\cdot 3 \cdot 4^{p-1}sp^2  }\Big(  \bigoplus_{i=1}^\Nsr \ket{\xbf_i}\bra{\xbf_i}^{\otimes p-1 } \otimes \Hbf_{\text{inho}}(\xbf_i)    \Big). $$
Finally, in order to remove the factor $\sqrt{\beta_{min}}/6 $, we can use  the amplification method~\cite{gilyen2019quantum} with further complexity $\mathcal{O}(6/\sqrt{\beta_{min}}  ) = \mathcal{O}(1)$ as it doesn't scale as any input parameter. In the end, we obtain the following operator:
$$ \frac{1}{4^{p-1} sp^2}\Big(  \bigoplus_{i=1}^\Nsr \ket{\xbf_i}\bra{\xbf_i}^{\otimes p-1 } \otimes \Hbf_{\text{inho}}(\xbf_i)    \Big),   $$
which has a similar form to Eqn.~(\ref{eqn: bigsumHessian}). The above construction was discussed for the case $C \geq 1$. For $C < 1$, all the steps are essentially the same (as in Section~\ref{sec: multipointhessian}), except that there will be an appearance of factor $C$ at the end, i.e., we would obtain the following operator: 
$$\frac{1}{(4C^2)^{p-1} sp^2}\Big(  \bigoplus_{i=1}^\Nsr \ket{\xbf_i}\bra{\xbf_i}^{\otimes p-1 } \otimes \Hbf_{\text{inho}}(\xbf_i)    \Big).    $$

Before moving further, we make a simple remark that for either case, $C \geq 1$ and $C < 1$, the operators of interest (e.g., the above one) only differ by a factor of $C^{2p-2}$. Now, we are ready to tackle our main objective, which is to dissect the convexity of a given function of arbitrary type. 

\subsection{Testing Positive-semidefiniteness}
\label{sec: testing}
Remind that we are interested in the spectrum, or more specifically, the minimum eigenvalue of Hessian $\Hbf$ at some given point $\xbf$. From Sec.~\ref{sec: overview} we have that $||\Hbf|| \leq p^2$ at any point $\xbf$, which means the eigenvalues of $\Hbf/p^2$ lie within $(-1,1)$. Given such a range of eigenvalues of $\Hbf$, the shifted matrix $(\Ibb_n - \Hbf/p^2)/2$ would have a spectrum lying in the range $(0,1)$. The reason why we change the attention to $(\Ibb_n - \Hbf/p^2)/2$ is because we can apply the improved quantum power method proposed in~\cite{nghiem2023improved} to find its maximum eigenvalue. 
\begin{lemma}[Theorem 2 of ~\cite{nghiem2023improved}]
\label{lemma: findingmin}
Given the block encoding of some positive-semidefinite matrix $A$ whose eigenvalues are $\in (0,1)$, then its largest eigenvalue can be estimated up to additive accuracy $\delta$ in time 
$$\mathcal{O}\Big( \frac{T_A}{\delta} \big( \log(\frac{1}{\delta}) + \frac{\log(n)}{2} \big)  \Big),$$
where $T_A$ is the complexity of producing the block encoding of $A$.
\end{lemma}

To see how it applies to our main problem, let $\lambda_{min}$ denotes the minimum eigenvalue of $\Hbf/p^2$. If $\lambda_{min} < 0$ (being negative) then $(1 - \lambda_{\min})/2 > 1/2$, and that $(1 - \lambda_{\min})/2$ is the maximum eigenvalue of $(\Ibb_n - \Hbf/p^2)/2$. Therefore, we can track sign of $\lambda_{min}$ from estimating minimum eigenvalue of $(\Ibb_n - \Hbf/p^2)/2$. \\

From the previous section we have obtained the block encoding of $\frac{1}{4^{p-1}} \bigoplus_{i=1}^\Nsr \ket{\xbf_i}\bra{\xbf_i}^{\otimes p-1 } \otimes \frac{ \Hbf(\xbf_i)}{2sp^2}$. The last recipe that we need to complete our algorithm is the following. First, suppose $\{A_i\}_{i=1}^M$ is a set of Hermitian operators, then the eigenspace of $\bigoplus_{i=1}^M A_i$ is simply the union of the eigenspace of all $\{A_i\}_{i=1}^M$~\cite{ccevik2011spectrum}. Second, the eigenspace of $\bigotimes_{i=1}^M A_i$ is the tensor product of the eigenspace of each $\{A_i\}$, which means that the eigenvalues of $\bigotimes_{i=1}^M A_i$ is the multiplication of eigenvalues of contributing matrices. From the second property, we can claim that for any $i$, the eigenvalues of $(1/4^{p-1}) \ket{\xbf_i}\bra{\xbf_i}^{\otimes p-1 } \otimes  \Hbf(\xbf_i)/(2sp^2) $ is the eigenvalues of $(1/4^{p-1}) \Hbf(\xbf_i)/(2sp^2)$. The reason is that, each operator $\ket{\xbf_i}\bra{\xbf_i}$ is a projector and hence, its only eigenvalue is $1$. Therefore, from the first property, we have that 
\begin{align*}
    \text{maximum eigenvalue of} \   \bigoplus_{i=1}^\Nsr \frac{1}{2} \frac{1}{2s 4^{p-1} } \big( \Ibb_n -  \ket{\xbf_i}\bra{\xbf_i}^{\otimes p-1 } \otimes \frac{\Hbf(\xbf_i)}{p^2} \big) = \max_{\xbf_i} \big\{ \frac{\Ibb_n - \Hbf(\xbf_i)/p^2}{4s 4^{p-1} }  \big\}_{i=1}^\Nsr.
\end{align*}

One may wonder why the right-hand side is important. We remind that if the function is convex in the given domain $\mathscr{D}$, then its Hessian matrix does not have non-negative eigenvalues at all points in $\mathscr{D}$. Therefore, if $\max_{\xbf_i} \{ \Hbf(\xbf_i)  \}$ is non-negative, then the Hessian is semi-definite at all $\Nsr$ points of consideration. As we pointed out from the beginning of this section, the condition of $\min_{\xbf_i} \{ \Hbf(\xbf_i)  \}$ being non-negative is equivalent to $\min_{\xbf_i} \{  (\Ibb_n - \Hbf(\xbf_i)) / 2 \}$ being no less than $1/2$. 

Fortunately, given the block encoding of $(1/4^{p-1})\bigoplus_{i=1}^\Nsr \ket{\xbf_i}\bra{\xbf_i}^{\otimes p-1 } \otimes  \Hbf(\xbf_i)/2sp^2$, it is simple to use lemma \ref{lemma: sumencoding} (plus a trivial block encoding of identity matrix of dimension $n\Nsr)$ to obtain the block encoding of their summation, which is exactly: 
$$ \bigoplus_{i=1}^\Nsr \frac{1}{4 sp^2 4^{p-1}} \big( \Ibb_n +  \ket{\xbf_i}\bra{\xbf_i}^{\otimes p-1 } \otimes \Hbf(\xbf_i) \big). $$
Then, we can find its minimum eigenvalue using recent work of~\cite{nghiem2023improved}, as we mentioned at the beginning of this subsection, e.g., Lemma~\ref{lemma: findingmin}. 

We finally remark that we are estimating the minimum eigenvalue of the operator 
$$ \bigoplus_{i=1}^\Nsr \frac{1}{4s 4^{p-1}} \big( \Ibb_n -  \ket{\xbf_i}\bra{\xbf_i}^{\otimes p-1 } \otimes  \frac{\Hbf(\xbf_i)}{p^2} \big), $$
which contains the extra factor $4s 4^{p-1}$. What we actually want is the maximum eigenvalue of the operator 
$$ \bigoplus_{i=1}^\Nsr \frac{1}{2} \big( \Ibb_n -  \ket{\xbf_i}\bra{\xbf_i}^{\otimes p-1 } \otimes\frac{\Hbf(\xbf_i) }{p^2}\big), $$
which corresponds exactly with 
$$ \max_{\xbf_i} \big\{ \frac{\Ibb_n - \Hbf(\xbf_i)}{2p^2}  \big\}_{i=1}^\Nsr. $$
Therefore, we need to use Lemma~\ref{lemma: findingmin} with an adjusted multiplicative error, e.g., by choosing
$$ \delta \longrightarrow  \frac{\delta}{2s 4^{p-1}}. $$

\smallskip \noindent {\bf Summary of the quantum algorithm procedure}. For convenience, we provide key points of our framework (which means we leave out technical details) plus corresponding complexity at each step.\\

$\bullet$ \ We begin with a set of points $\{\xbf_i\}_{i=1}^\Nsr$ (each $\xbf_i \in \mathbb{R}^n$ and $|\xbf_i| \leq 1$) of interest. Define $C^2 = \sum_{i=1}^\Nsr |\xbf_i|^2$ as normalization factor. \\

$\bullet$ \ Load the above points into quantum state $\ket{\phi} = \frac{1}{C} \sum_{i=1}^\Nsr \ket{i} \otimes \xbf_i$, then construct the block encoding of the following operator
\begin{align*}
    \frac{1}{C^2}\begin{pmatrix}
        \xbf_1\xbf_1^T & \cdot & \cdot &\cdot \\
        \cdot & \xbf_2 \xbf_2^T & \cdot & \cdot \\
        \cdot & \cdot & \cdot & \cdot & \\
        \cdot & \cdot & \cdot & \xbf_\Nsr \xbf_\Nsr^T 
    \end{pmatrix}.
\end{align*}
The complexity of the above step is $\mathcal{O}( \log(n\Nsr )$.\\

$\bullet$ \ Break into two cases $C \geq 1$ and $C < 1$. First, consider $C \geq 1$, then use amplification~\cite{gilyen2019quantum} to remove the factor $C^2$. The complexity for amplification step is $\mathcal{O}(C^2 \log(n\Nsr))$. Then we 
construct the following operator 
\begin{align*}
    \begin{pmatrix}
        0.5 |\xbf_1|  \ket{\xbf_1}\bra{\xbf_1} & \cdot & \cdot &\cdot \\
        \cdot & 0.5 |\xbf_2| \ket{\xbf_2}\bra{\xbf_2} & \cdot & \cdot \\
        \cdot & \cdot & \cdot & \cdot & \\
        \cdot & \cdot & \cdot & 0.5 |\xbf_\Nsr| \ket{\xbf_\Nsr} \bra{\xbf_\Nsr} 
    \end{pmatrix}.
\end{align*}
The complexity of this step is $\mathcal{O}( \frac{C^2}{\xbf_{min}} \log(n\Nsr)   )$ where $\xbf_{min} \equiv \min \{ |\xbf_i|^2 \}_{i=1}^\Nsr$.\\

$\bullet$ \ Using the above operator, we construct the following operator 
\begin{align*}
\frac{1}{2^{p-1}} \begin{pmatrix}
\big( |\xbf_1| |\xbf_1 \rangle \langle \xbf_1 |  \big)^{\otimes p-1} \otimes \Ibb_n  & & & \\
& \big( |\xbf_2| |\xbf_2 \rangle \langle \xbf_2 |  \big)^{\otimes p-1} \otimes \Ibb_n & & \\
& & \cdots& \\
& & & \big( |\xbf_\Nsr| |\xbf_\Nsr \rangle \langle \xbf_\Nsr |  \big)^{\otimes p-1} \otimes \Ibb_n
\end{pmatrix}.
\end{align*}
The complexity upon this step is $\mathcal{O}( p \frac{C^2}{\xbf_{min}} \log(n\Nsr)  )$. \\

$\bullet$ Use oracle access to $A$ (defined in section \ref{sec: overview}) to construct the block encoding of $(M_H + M_D)/2sp^2$. The complexity of this step is $\mathcal{O}( p^2 \log(n)  )$. \\

$\bullet$ \ Employing mathematical property of homogeneous polynomial (see \ref{sec: singlepointhessian}), we construct the block encoding of 
$$ \frac{1}{4^{p-1}} \bigoplus_{i=1}^\Nsr (\ket{\xbf_i}\bra{\xbf_i})^{\otimes p-1} \otimes \frac{\Hbf(\xbf_i)}{2sp^2}  $$
that contains our Hessian $\Hbf$ (for homogeneous polynomial). \\

$\bullet$ \ Generalize the above construction to a polynomial of arbitrary type. The complexity of this step (including everything from the beginning) is
$$ \mathcal{O}\big( \frac{C^2}{\xbf_{min}} \log(n\Nsr)  + p^2 \log(n) \big). $$

$\bullet$ \ Construct the block encoding of the operator 
$$ \bigoplus_{i=1}^\Nsr \frac{1}{4 s 4^{p-1}} \big( \Ibb_n -  \ket{\xbf_i}\bra{\xbf_i}^{\otimes p-1 } \otimes \frac{\Hbf(\xbf_i}{p^2}) \big).  $$

$\bullet$ \ Find minimum eigenvalue of the above operator by using Lemma~\ref{lemma: findingmin} (with accuracy being scaled $\epsilon/4sp^2 4^{p-1}$, and infer the convexity from such eigenvalue. More specifically, if the minimum eigenvalue is less than $1/2$, then the function is not convex. Otherwise, it is convex. The complexity of this step is:
$$ \mathcal{O} \Big( 4^{p-1} sp^2 ( \log(n\Nsr) + \log( 4^{p-1} sp^2 ) ) ( p \frac{C^2}{\xbf_{min}} (\log(n\Nsr) + p^2 \log(n) ) \Big).  $$

$\bullet$ \ Finally consider the case where the normalization factor $C < 1$. Repeat the same procedure, which results in almost the same complexity, i.e.,
$$ \mathcal{O} \Big( (4C^2)^{p-1} sp^2 ( \log(n\Nsr) + \log( 4^{p-1} sp^2 ) ) ( p \frac{C^2}{\xbf_{min}} (\log(n\Nsr) + p^2 \log(n) ) \Big).  $$

We remark that in the above summary, we did not take into account the error term $\epsilon$ that appears in multiple steps, such as amplification, encoding matrices $M_H,M_D$, and taking the square root of the operator. To dissect the convexity, we can ignore the error-dependence factors because we only care about the sign of the minimum eigenvalue instead of a real estimation. Therefore, one can set the error to be some constant. Furthermore, the complexity dependence on the error is polylogarithmic, which is efficient. The generalization to polynomials of arbitrary type is carried out in Section~\ref{sec: generalization}. A subtle detail about the above running time is that the scaling depends on $|\xbf_{min}|$, which is the minimum norm of the length of chosen points, as well as $C$, which is the normalization factor. Let's say we choose $\Nsr$ points to be distributed uniformly in the domain of interest (hypersphere), then we can assume that $|\xbf_{min}| \sim \mathcal{O}(1)$ (being some constant) and hence $1 \leq C^2 \leq \Nsr$. So the factor $C^2/|\xbf_{min}| \in \mathcal{O}(\Nsr)$, which is linear in $\Nsr$. For $C$ being smaller than $1$, which means that we choose $\Nsr$ points having very small norm (at least, for most of them). Then we can assume that $|\xbf_{min}|$ is smaller than $1/\Nsr$, which means that $1/|\xbf_{min}|$ is greater than $\Nsr$, but no greater than $\mathcal{O}(\Nsr)$. All in all, for both cases where $C \geq 1$ or $C < 1$, the factor $C^2/|\xbf_{min}|$  is $\mathcal{O}(\Nsr)$, which is linear in $\Nsr$.  \\

Now we are ready to state our main result formally:
\begin{theorem}[Testing Convexity over $\Nsr$ Sample Points]
\label{thm: main}
    Let an objective function $f$ (of arbitrary type), with certain assumptions defined as in Sections~\ref{sec: overview} and~\ref{sec: generalization}. Over the domain $\mathscr{D} \subseteq (-1,1)^n$, let $\mathscr{N}$ be the chosen sample points, denoted as $\{ \xbf_i \}_{i=1}^\Nsr$ for convexity testing. Let $C^2 = \sum_{i=1}^\Nsr |\xbf_i|^2 (\leq \Nsr))$  and $\xbf_{min} = \min_i \{ |\xbf_i|^2 \}_{i=1}^\Nsr$. If $C \geq 1$, then the quantum algorithm outlined above can reveal the positive-semidefiniteness of corresponding Hessian in time 
    $$ \mathcal{O}\Big( 4^{p-1}  sp^2  \ ( \log(n\Nsr) + \log(4^{p-1} 2sp^2)) \ ( p \Nsr (\log(n\Nsr)) +  p^2 \log(n) \Big), $$
    where $\alpha$ is some bounded constant. In particular, if $C < 1$, then the running time is: 
    $$ \mathcal{O}\Big( (4C^2)^{p-1} sp^2  \ (\log(n\Nsr) + \log(4^{p-1} 2sp^2)) \ ( p \Nsr (\log(n\Nsr)) +  p^2 \log(n) \Big). $$
\end{theorem}

\smallskip \noindent {\bf Potential advantage}\\
To see the advantage, we take a look at how classical computers can solve the above problem. The Hessian matrix is computed as
\begin{align}
    \Hbf(\xbf) = \Tr_{1,2,...,p-1} \big( (M_H + M_D)  \ ( \rho_x^{\otimes p-1} \otimes \Ibb_n )   \big),
\end{align}
which involves matrix operation of size $n^p$. Each operator $M_H$ and $M_D$ is computed from matrix $A$ which takes further time $\mathcal{O}( pn^p + p^2 n^p )$. The Hessian is then evaluated at $\Nsr$ points to find the sign of minimum eigenvalue, which results in  the final complexity $\mathcal{O}(\Nsr ( pn^p + p^2 n^p))$.

The above running time clearly suggests that the quantum algorithm can test convexity superpolynomially more efficiently than its classical counterpart with respect to the number of variables $n$, meanwhile a bit slower in relative to the number of sample points $\Nsr$. Therefore, the quantum method is very efficient when dealing with high-dimensional cases as well as high-degree polynomials.\\

An interesting question from the above classical procedure is: what if we follow the same routine using a quantum computer for all points? It means that one can run the quantum framework to check the spectrum of Hessian at each point $\xbf$, and repeat the procedure for $\Nsr$ points. In such a case, the quantum procedure would be very similar to our prior construction, except that we do not care about other $\Nsr - 1$ points, more specifically, if we take a look at Eqn.~\ref{eqn: C} and pay attention only to the top-left corner, e.g., the operator $\xbf_1 \xbf_1^T$ (the factor $C$ is treated in a similar manner as the above construction), which means that we treat this point $\xbf_1$ as a point of interest. Then, the same procedure can be carried out to first build the squared operator:
\begin{align}
    \begin{pmatrix}
        \xbf_1\xbf_1^T & \cdot & \cdot &\cdot \\
        \cdot & \cdot & \cdot & \cdot \\
        \cdot & \cdot & \cdot & \cdot & \\
        \cdot & \cdot & \cdot & \cdot
    \end{pmatrix} \longrightarrow  
    \begin{pmatrix}
        0.5 |\xbf_1|  \ket{\xbf_1}\bra{\xbf_1} & \cdot & \cdot &\cdot \\
        \cdot & \cdot & \cdot & \cdot \\
        \cdot & \cdot & \cdot & \cdot & \\
        \cdot & \cdot & \cdot & \cdot  
    \end{pmatrix}.
\end{align}
Then, from the right-handed side operator, one proceeds to build the block encoding of $\Big( 0.5 |\xbf_1|\ket{\xbf_1}\bra{\xbf_1}  \Big)^{\otimes p-1} \otimes \Ibb_n$ (using the routine below Equation~(\ref{eqn: squared})). Then, one uses the same properties of Hessian (see Eqn.~(\ref{eqn: hessian})) to obtain the following:
\begin{align}
    \Big( 0.5 |\xbf_1|\ket{\xbf_1}\bra{\xbf_1}  \Big)^{\otimes p-1} \otimes \Ibb_n  \cdot \ \frac{(M_H+M_D)}{2sp^2} \cdot \Big( 0.5 |\xbf_1|\ket{\xbf_1}\bra{\xbf_1}  \Big)^{\otimes p-1} \otimes \Ibb_n = \frac{1}{4^{p-1}} \ket{\xbf_1}\bra{\xbf_1}^{p-1} \otimes \frac{\Hbf(\xbf_1)}{2sp^2}.
\end{align}
One can see that the above formula is essentially the top-left block corner of Eqn.~(\ref{eqn: bigsumHessian}), which is quite obvious. Then, one can use the same Lemma~\ref{lemma: findingmin} to reveal the minimum value, which indicates the positivity of Hessian $\Hbf$ at given point $\xbf_1$. One repeats the process for $\Nsr$ different points, which results in asymptotically the same complexity. The difference is that, as we need to store $\Nsr$ different eigenvalues (of Hessian at $\Nsr$ points), the memory usage is required to be as much as $\Nsr$. Meanwhile, the quantum process that we have outlined during this work requires $\mathcal{O}(\log(\Nsr))$ qubits to handle the same task, which is more effective in terms of memory usage but sharing the same running time. 
\section{Remarks and Discussion}
\label{sec: remarkanddiscussion}

In this section, we discuss our algorithm in a larger context, showing the potential application of our result in multiple directions. 

\subsubsection{From Hessian to Curvature and Geometric Structure of Manifold}
\begin{figure}[H]
    \centering
    \begin{tikzpicture}[scale=1.5]
    
    \coordinate (A) at (2.0,0,0);
    \coordinate (B) at (0,2.0,0);
    \coordinate (C) at (0,0,2.0);

    \draw[fill=blue!20,opacity=0.5] (A) -- (B) -- (C) -- cycle;

    \draw[->] (0,0,0) -- (2.5,0,0) node[right]{$y$};
    \draw[->] (0,0,0) -- (0,2.5,0) node[above]{$z$};
    \draw[->] (0,0,0) -- (0,0,2.5) node[below left]{$x$};

    \node[above right] at (A) {$A$};
    \node[above right] at (B) {$B$};
    \node[below right] at (C) {$C$};
    \node[above right] at (1,1,0) {Plane};
\end{tikzpicture}
\hspace{1cm} 
\begin{tikzpicture}[scale=1.5]
    
    \coordinate (O) at (0,0,0);
    \def\r{1.2}

    \draw[ball color=blue,opacity=0.3] (O) circle (\r);

    \draw[->] (0,0,0) -- (2.5,0,0) node[right]{$y$};
    \draw[->] (0,0,0) -- (0,2.5,0) node[above]{$z$};
    \draw[->] (0,0,0) -- (0,0,2.5) node[below left]{$x$};

    \node[above right] at (O) {$O$};
    \node[above right] at (1,1,0) {Sphere};
\end{tikzpicture}
    \caption{A plane and a sphere. A plane has a canonical representation as $ax+by+cz = d$ where $a,b,c,d$ are parameters that defined the plane. Thus, a point on the plane has $z$-coordinate $z = -(a/c)x - (b/c)y + d$, which could be treated as a function of two variables. On the other hand, a sphere is characterized by the equation $x^2+y^2+z^2 =r^2$ where $r$ is the radius. Equivalently, $z =\pm \sqrt{r^2 - x^2 -y^2}$. }
    \label{fig: planesphere}
\end{figure}
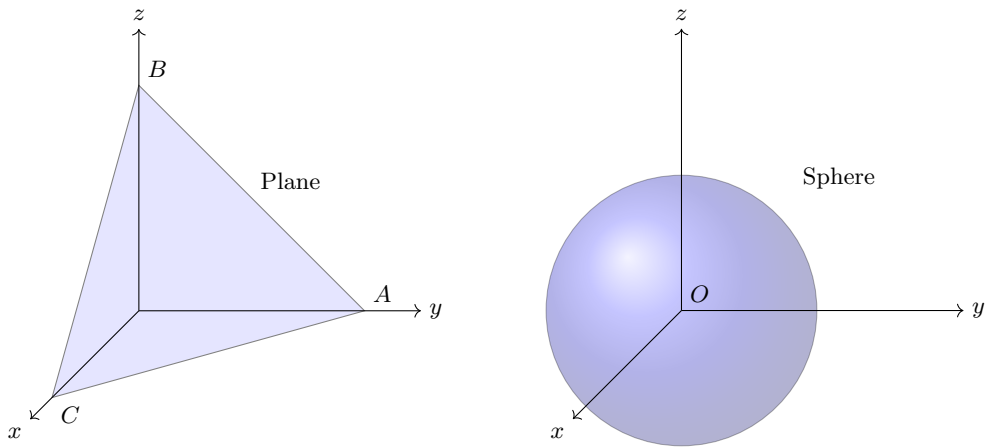

As we mentioned, the Hessian of a function at a given point encodes the local geometric structures of such a function, e.g., convexity. While the sign of eigenvalues of Hessian can reveal the convexity, the magnitude of eigenvalues can actually reveal how curved it is (see Fig.~\ref{fig: planesphere} for a simple example in 3D). Figure \ref{fig: planesphere} features two simple surfaces in 3D, which are a plane and a sphere. For the plane, it is easy to compute the Hessian of $z(x,y) = -(a/c)x - (b/c)y + d/c$ in this case, which is simply a $2 \times 2$ matrix of 0 entries, which implies that the surface is not curved at all, i.e., exactly match the plane shape. On the other hand, for a sphere, $z =\sqrt{r^2 - x^2 -y^2}$ so the Hessian matrix is not as straightforward as a plane, but as we can see, the function $z$ is convex on the lower half of plane $x-y$, but not on the upper half. 

In fact, studying the underlying structure of a manifold using functions defined on the manifold is a fundamental aspect of differential geometry. Thus, from this angle, our work suggests a potential aid of a quantum computer in the study of the geometric structure of a manifold, given that a manifold is locally Euclidean and arbitrary smooth function of some variables can be approximated by certain polynomials in some domains.

\subsubsection{Variational Quantum Algorithm}
A popular topic of recent progress in quantum computation is the variational quantum algorithm (VQA)~\cite{cerezo2021variational, mcclean2016theory, havlivcek2019supervised, mitarai2018quantum, peruzzo2014variational}. One of the reasons that make variational quantum algorithms so appealing is that typically, they require low-depth circuits, which is very suitable in the near-term era. VQA has shown its success in many areas such as combinatorial optimization \cite{farhi2014quantum}, supervised learning \cite{mitarai2018quantum}, etc. A common strategy in the context of VQA eventually boils down to the minimization problem: 
$$ \text{min}_{\theta} \ f(\theta) = \bra{\textbf{0}} U(\theta)^\dagger O U(\theta) \ket{\textbf{0}}, $$
where $U(\theta)$ refers to some variational circuit, e.g., a circuit composed of rotational gates with adjustable angles. A common method to optimize the above quantity is gradient descent, where the above observables are usually defined as a cost function, and its gradient is computed classically. Then, the parameters are updated iteratively by tuning corresponding rotational gates. Optimizing a quantum circuit is apparently not easy, as a phenomenon called the barren plateau can occur~\cite{wang2021noise, cerezo2021cost}, which prevents the efficient training of quantum circuits. In particular, training general variational quantum circuits is even NP-hard~\cite{bittel2021training}.

Theoretically, the cost landscape of the above function is also a factor that affects the minimization. If the domain is convex, then any initial randomization falling into that domain can lead to a minimum, which implies that the optimization is efficient. While the exact expression for $f(\theta)$ might not necessarily be a polynomial, we remark that over some domains that are sufficiently small, arbitrary functions can be approximated by some polynomials, e.g., Taylor series. Therefore, a visible strategy that we see from our work is that we can consider multiple domains and check the convexity. Given that the task of dissecting convexity can be done superpolynomially fast in the quantum realm, it turns out to be quite useful in this case as we can scan the optimization landscape efficiently with respect to the number of parameters. The challenge then only arises as to how to represent $f(\theta)$ with proper polynomials, which allows computable parameters in a form that we assume in our work. At first glance, this challenge seems impossible for a random circuit $U$, but might not be so for a structured $U$ that has been employed in several contexts, such as the so-called QAOA ansatz~\cite{lloyd2020quantum}. We leave this question as a motivation for future exploration of the application of our convexity tester. 

\subsubsection{Improving Quantum Newton's Method}
As a striking corollary of our method is the generalization to arbitrary polynomial as well as major improvement on the resource requirement, upon the quantum Newton's method developed in~\cite{rebentrost2019quantum}. In~\cite{nghiem2023improved}, we already made progress on quantum gradient descent, and in fact, the method outlined in this work shares certain similarities with what in~\cite{nghiem2023improved}, but the scope is different. 

To begin with, we recall some aspects of quantum Newton's method, which is basically a modified gradient descent method. We first follow the same notation and assumption from~\cite{rebentrost2019quantum}. In such a problem, we are given an objective function $f: \mathbb{R}^n \longrightarrow \mathbb{R}$, which is a homogeneous polynomial of even degree as we defined in Section~\ref{sec: overview}. The goal is to find the point $\xbf \in \mathbb{R}^n$ at which $f(\xbf)$ is minimum. A standard method for this kind of optimization problem is gradient descent, which is an iterative method in that one begins with a random solution $\xbf_0$ and performs the update iteratively as follows:
\begin{align}
    \xbf_{t+1} = \xbf_t - \eta \nabla f(\xbf_t). 
\end{align}
As mentioned in Section~\ref{sec: overview}, such particular form of $f$ admits an analytical form for the gradient, as $\nabla f(\xbf) = D(\xbf) \xbf$, which was used in~\cite{rebentrost2019quantum} (and improved in~\cite{nghiem2023improved}) to construct the quantum process carrying out the above iteration procedure. As also mentioned in the same work~\cite{rebentrost2019quantum}, Newton's method modifies directly upon the gradient descent method by taking account of the curvature of $f$, i.e., the update rule is as follows:
\begin{align}
     \xbf_{t+1} = \xbf_t -  \eta \ \Hbf^{-1}(\xbf_t) \nabla f(\xbf_t). 
\end{align}
Roughly speaking, at a given time step $t$-th, the method in \cite{rebentrost2014quantum} takes multiple copies of $\xbf_{t}$, use oracle access to simulate $\exp(-iM_D t)$ and $\exp(-i M_H t)$, to construct the gradient and Hessian (via relation in Equation \ref{eqn: 6} and \ref{eqn: 10}). Then they perform the subtraction of vectors by using extra ancilla and Hadamard gates, to obtain $\xbf_{t+1}$. If one wish to obtain a normalized version of temporal solution $\xbf_{t+1}$, $\ket{\xbf}_{t+1}$, then by measuring the ancilla and post-select on ancilla being $\ket{0}$, one achieves the goal, as in \cite{rebentrost2014quantum}. Here, we lift such requirement and consider the problem in a more general manner, that is obtaining a temporal solution written in a form similar to ``density operator'', $\xbf_t \xbf_t^T$ for any given time step $t$.    \\

From Section~\ref{sec: singlepointhessian}, \ref{sec: multipointhessian} (particularly equation~(\ref{eqn: singlehessian})) and most importantly section \ref{sec: generalization}, we have the block encoding of $(\xbf \xbf^T)^{\otimes p-1} \otimes \Hbf(\xbf)/(2sp^2)$ where $\Hbf$ refers to the Hessian of polynomial of arbitrary kind. What is lacked is the gradient of polynomial of arbitrary kind, which was not constructed before and neither in the previous work \cite{nghiem2023improved}, so here we first fill this gap. Recall that from section \ref{sec: generalization}, we have that for a general polynomial:
\begin{align}
    f(\xbf) = \sum_{q=1}^{P-1} (c_q^T \xbf) \  \prod_{k=1}^{q-1} (\xbf^T B_{k q } \xbf).
\end{align}
Since the derivative of a sum is equal to sum of derivative of each term within the summation, then for simplicity, as similar to what we did in section \ref{sec: generalization}, we consider each constituent of the above summation, which has the form $g(\xbf)  =  (c^T \xbf) \  \prod (\xbf^T B \xbf) \equiv (c^T \xbf) h(\xbf)$ (where we have defined $\prod (\xbf^T B \xbf) \equiv h(\xbf) $. Then by chain rule, it is simple to see that:
\begin{align}
    \frac{\partial g}{\partial x_m} &= (c^T\xbf) \frac{\partial h(\xbf)}{\partial x_m}  + h(\xbf) \frac{\partial c^T \xbf }{\partial x_m} \\
    &= (c^T\xbf) \frac{\partial h(\xbf)}{\partial x_m} + h(\xbf) \frac{\partial \sum_{i=1}^n c_j x_j }{\partial x_m} \\
    &= (c^T\xbf) \frac{\partial h(\xbf)}{\partial x_m}  + h(\xbf) c_m
\end{align}
where $x_m$ is the $m$-th variable of $\xbf$ and $c_m$ is the $m-th$ entry of $c$ (note that $1 \leq m \leq n$ and $n$ is th dimension of $\xbf$). As the gradient of a function is composed of partial derivative of such function with respect to all variables, we have that:
\begin{align}
    \Vec{\nabla} g(\xbf) = (c^T \xbf) \Vec{\nabla} h(\xbf) + h(\xbf) c
\end{align}
Since $h(\xbf)$ is a homogeneous even degree polynomial, its gradient admits an explicit analytical expression. Recall that from section \ref{sec: singlepointhessian} and property in equation \ref{eqn: 6}, we have that:
\begin{align}
    (\xbf\xbf^T) ^{\otimes p-1} \otimes \Ibb_n \ ( M_D) \ (\xbf\xbf^T) ^{\otimes p-1} \otimes \Ibb_n &= (\xbf\xbf^T) ^{\otimes p-1} \otimes D(h(\xbf))
\end{align}
where $D(h(\xbf))$ refers directly to the fact that it is the gradient operator of $h(\xbf)$ evaluated at the point $\xbf$. 

Recall that via oracle access to $A$, we have $\epsilon$-approximated block encoding of $M_D/(sp)$. Then we can use lemma \ref{lemma: product} to construct the block encoding of 
\begin{align}
(\xbf\xbf^T) ^{\otimes p-1} \otimes \Ibb_n \ \frac{M_D}{sp} \ (\xbf\xbf^T) ^{\otimes p-1} \otimes \Ibb_n = (\xbf\xbf^T) ^{\otimes p-1} \otimes \frac{D(h(\xbf))}{sp} 
\end{align}
We also have block encoding of $cc^T$ (by assumption, see section \ref{sec: generalization}). Lemma \ref{lemma: product} allows us to construct the block encoding of $(\xbf\xbf^T) \ (cc^T) = (\xbf^T c)\xbf c^T$. Then using lemma \ref{lemma: tensorproduct} using block encoding of $\Ibb$, we can obtain block encoding of $ \Ibb \otimes (\xbf^T c)\xbf c^T$. Then lemma \ref{lemma: product} allows us to construct the block encoding of 
\begin{align}
(\xbf\xbf^T) ^{\otimes p-1} \otimes \frac{D(h(\xbf)) (\xbf^T c)\xbf c^T }{sp} = (\xbf\xbf^T) ^{\otimes p-1} \otimes \frac{ (\xbf^T c) \Vec{\nabla} h(\xbf) c^T}{sp}  
\end{align}
Using lemma \ref{lemma: product} again with block encoding of $\Ibb \otimes (c^T\xbf) c\xbf^T$ (which is the transpose of $(\xbf c^T) \xbf c^T$), we obtain the block encoding of 
\begin{align}
(\xbf\xbf^T) ^{\otimes p-1} \otimes \frac{ (\xbf^T c) \Vec{\nabla} h(\xbf) c^T \ (c^T\xbf) c\xbf^T  }{sp}  =    (\xbf\xbf^T) ^{\otimes p-1} \otimes \frac{ (\xbf c^T)^2 \Vec{\nabla} h(\xbf) \xbf^T  }{sp}  
\end{align}
where we have use $c^T c = 1$ and $\xbf^T c = c^T \xbf$ due to the real regime that we work on. Now we handle the term $h(\xbf) c$. Since $h(\xbf)$ is homogeneous even degree, it has the familiar form:
\begin{align}
    h(\xbf) = \frac{1}{2} \bra{\xbf}^{\otimes p} A \ket{\xbf}^{\otimes p}
\end{align}
As we have oracle access to $A$, we can construct the block encoding of $A/2s$. From block encoding of $\xbf \xbf^T$, it is trivial to obtain the block encoding of $(\xbf \xbf^T)^{\otimes p}$ using lemma \ref{lemma: tensorproduct}. We have that:
\begin{align}
    (\xbf \xbf^T)^{\otimes p} \frac{A}{2s} (\xbf \xbf^T)^{\otimes p} = (\xbf \xbf^T)^{\otimes p-1} \otimes \frac{h(\xbf) }{s}\xbf \xbf^T 
\end{align}
Now we use the block encoding of $\Ibb \otimes cc^T$ plus lemma \ref{lemma: product} to construct the block encoding of
\begin{align}
   ( \Ibb \otimes cc^T) \ ( (\xbf \xbf^T)^{\otimes p-1} \otimes \frac{h(\xbf) }{s}\xbf \xbf^T   ) = (\xbf \xbf^T)^{\otimes p-1} \otimes \frac{ (c^T\xbf) h(\xbf) c\xbf^T   } {s}
\end{align}
We use lemma \ref{lemma: scale} to add a scaling of $1/p$ to the above term, e.g, obtaining the block encoding of $(\xbf \xbf^T)^{\otimes p-1} \otimes \frac{ (c^T\xbf) h(\xbf) c\xbf^T   } {ps}$. Then we use lemma \ref{lemma: sumencoding} to construct the block encoding of:
\begin{align}
  &  \frac{1}{2} \big(  (\xbf\xbf^T) ^{\otimes p-1} \otimes \frac{ (\xbf c^T)^2 \Vec{\nabla} h(\xbf) \xbf^T  }{sp}  +(\xbf \xbf^T)^{\otimes p-1} \otimes \frac{ (c^T\xbf) h(\xbf) c\xbf^T   } {ps}  \big) \\
   &= \frac{1}{2} \big(  (\xbf \xbf^T)^{\otimes p-1} \otimes \frac{\xbf^T c}{sp} ( (\xbf^T c) \Vec{\nabla} h(\xbf)\xbf^T + h(\xbf)c \xbf^T  )   \big)  \\
   &=  \frac{1}{2} \big(  (\xbf \xbf^T)^{\otimes p-1} \otimes \frac{\xbf^T c}{sp} \  \Vec{\nabla} g(\xbf)\xbf^T \big) 
\end{align}
Before moving further, we remind that we have the ($\epsilon$-approximated) block encoding of the following operator:
$$ (\xbf \xbf^T )^{\otimes p-1} \otimes \frac{\Hbf(\xbf)}{2sp^2} \ \text{ and }  \  \frac{1}{2} \big(  (\xbf \xbf^T)^{\otimes p-1} \otimes \frac{\xbf^T c}{sp} \  \Vec{\nabla} g(\xbf)\xbf^T \big) $$

Now we have enough recipe to deal with quantum Newton's method. Let us recall a lemma from~\cite{nghiem2023improved} (note that in their context, $xx^T$ is exactly $\xbf \xbf^T$ in our case):
\begin{lemma}[Lemma 11 in~\cite{nghiem2023improved}]
     Given the block encoding of $(x x^T)^{\otimes p-1} \otimes \frac{D(x)}{ps}$ (with complexity $T$), then it is possible to obtain the block encoding of $D(x)/p$ in time $\mathcal{O}( \gamma^{2(p-1)} s T_D )$, where $\gamma$ is some bounded constant.
\end{lemma}
The above form coincides with what we have right above. Therefore, using the same procedure, we can obtain the block encoding of $\Hbf(\xbf)/p^2$ in similar time 
$$\mathcal{O}\Big(\gamma^{2p-2} s \ \big( p^2\log(n) + p\log^{2.5}(1/\epsilon)\big)\Big).$$ 
and we also obtain the block encoding of $(\xbf^T c) \Vec{\nabla} g(\xbf)\xbf^T/p $ in time
$$ \mathcal{O}(  \gamma^{2(p-1)}  s \big( p^2\log(n) + p\log^{2.5}(1/\epsilon)\big)\Big).    ) $$
The extra factor $x^T c$ can be estimated up to small accuracy by using routine in appendix \ref{sec: alphadiagonal}, setting $\Nsr = 1$ and using amplitude estimation. The concrete procedure will be elaborated in appendix \ref{sec: alphadiagonal}. Then it is removed from the above operator using amplification method, with further $\mathcal{O}( 1/(x^T c)) = \mathcal{O}(1)$ complexity. Therefore, we obtain the $\epsilon$-approximated block encoding of $\Vec{\nabla} g(\xbf)\xbf^T/p $ in complexity 
$$  \mathcal{O}(  \gamma^{2(p-1)}  s \big( p^2\log(n) + p\log^{2.5}(1/\epsilon)\big)\Big) $$

The minimum eigenvalue (in magnitude) of $\Hbf (\xbf)/p^2$ can be found exactly via the convexity testing framework (see section \ref{sec: generalization} and \ref{sec: testing}), which is required to perform inversion of $\Hbf(\xbf)$ in order to execute quantum Newton's method. The inversion of $\Hbf(\xbf)/p^2$, or more precisely, the transformation from $\Hbf(\xbf)/p^2 \longrightarrow \Hbf^{-1}(\xbf)/\Gamma$, where $\Gamma$ is roughly $\kappa$, where $\kappa$ is roughly a reciprocal of minimum eigenvalue of $\Hbf(\xbf)$ (according to~\cite{childs2017quantum}), is then carried out using popular methods~\cite{childs2017quantum, gilyen2019quantum}, with running time
$$  \mathcal{O}\Big(\gamma^{2p-2} 2s \ \big( p^2\log(n) + p\log^{2.5}(\frac{1}{\epsilon}) \big) \  \kappa \  polylog(\frac{\kappa}{\epsilon}) \Big), $$
Therefore, it is sufficient to complete the improved quantum Newton's method. More specifically, at $t$-th time step, we are presented with $\xbf_t \xbf_t^T$. In particular, $\xbf_t \xbf_t^T$ is related to $\xbf_{t+1} \xbf_{t+1}^T$ via the following relation, that was used in \cite{nghiem2023improved}:
\begin{align}
   \xbf_{t+1} (\xbf_{t+1})^T &= \big (\xbf_t  - \eta \Hbf^{-1}(\xbf_t) \Vec{\nabla} g(\xbf_t)\big)\big( \xbf_t  - \eta \Hbf^{-1}(\xbf_t) \Vec{\nabla} g(\xbf_t)  \big)^T \\
    &=  \big (\xbf_t  - \eta \Hbf^{-1}(\xbf_t) \Vec{\nabla} g(\xbf_t)\big) \big(  \xbf_t^T - \eta(  \Hbf^{-1}(\xbf) \Vec{\nabla} g(\xbf_t))^T     \big) \\
    &= (\xbf_t \xbf_t^T) - \eta \xbf_t ( \ \Hbf^{-1}(\xbf_t) \Vec{\nabla} g(\xbf_t) \ )^T  - \eta \Hbf^{-1}(\xbf_t) \Vec{\nabla} g(\xbf_t) \xbf_t^T + \\  
    &  \eta^2 (\Hbf^{-1}(\xbf_t) \Vec{\nabla} g(\xbf_t)  )  \cdot (\Hbf^{-1}(\xbf_t) \Vec{\nabla} g(\xbf_t)  )^T  \\
\end{align}
The block encoding of $\xbf_t \xbf_t^T$ is apparently presented, more concretely, in quantum Newtons' method, it is from the previous $t-1$-th step. Previously, we have constructed the ($\epsilon$-approximated ) block encoding of $H^{-1}/\Gamma$ and of $\Vec{\nabla} g(\xbf_t) \xbf_t^T$. The block encoding of $\Vec{\nabla} \frac{1}{p} g(\xbf_t) \xbf_t^T$ naturally yields the block encoding of $(\Vec{\nabla} \frac{1}{p} g(\xbf_t) \xbf_t^T)^T = \frac{1}{p}\xbf_t \Vec{\nabla} g(\xbf_t)^T$. Then we can use lemma \ref{lemma: product} to construct the block encoding of 
   $$\frac{1}{p} \xbf_t \ \Vec{\nabla} g(\xbf_t)^T \ (\frac{\Hbf^{-1}(\xbf_t)}{\Gamma})^T $$
We then can use lemma \ref{lemma: scale} to insert the factor $\eta$, that we transform the block encoding of the above operator into 
$$ \frac{\eta}{p} \xbf_t \ \Vec{\nabla} g(\xbf_t)^T \ (\frac{\Hbf^{-1}(\xbf_t)}{\Gamma})^T  $$
The unitary transpose of the block encoding of above operator is exactly  the block encoding of
$$  \frac{\eta}{p} \frac{\Hbf^{-1}(\xbf_t)}{\Gamma} \ \Vec{\nabla} g(\xbf_t)  \xbf_t^T $$
We can use lemma \ref{lemma: product} to construct the block encoding of their product, e.g.,
$$ \big( \frac{\eta}{p} \frac{\Hbf^{-1}(\xbf_t)}{\Gamma} \ \Vec{\nabla} g(\xbf_t)  \xbf_t^T \big) \cdot \big(  \xbf_t \ \Vec{\nabla} g(\xbf_t)^T \ (\frac{\Hbf^{-1}(\xbf_t)}{\Gamma})^T \frac{\eta}{p} \big) \\ 
   = |\xbf_t|^2 \frac{\eta^2}{p^2} \frac{1}{\Gamma^2} (\Hbf^{-1}(\xbf_t) \Vec{\nabla} g(\xbf_t)  )  \cdot (\Hbf^{-1}(\xbf_t) \Vec{\nabla} g(\xbf_t)  )^T   $$
In order to remove the factor $|\xbf_t|^2$, we just need to note that since we have the block encoding of $\xbf_t \xbf_t^T$, we can use lemma \ref{lemma: findingmin} to efficiently find its maximum eigenvalue, which is exactly $|\xbf_t|^2$. Then we can use the amplification method to remove such factor, resulting in further complexity $\mathcal{O}( 1/|\xbf_t|^2 )$. 

Due to the term $p^2 \Gamma^2$ (which is known as roughly one over minimum eigenvalue of $\Hbf(\xbf_t)$ , we need to use lemma \ref{lemma: scale}) to change the block encoding of $\xbf_t \xbf_t^T $ into $\frac{1}{p^2\Gamma^2} \xbf_t \xbf_t^T$,  and of $(\eta/p) \xbf_t \ \Vec{\nabla} g(\xbf_t)^T \ (\frac{\Hbf^{-1}(\xbf_t)}{p \Gamma})^T  $ into\  $\eta \xbf_t \ \Vec{\nabla} g(\xbf_t)^T \ (\frac{\Hbf^{-1}(\xbf_t)}{p \Gamma^2})^T $ (same thing for its transpose). Then we employ lemma \ref{lemma: sumencoding} to construct the block encoding of 
$$ \frac{1}{4p^2 \Gamma^2}((\xbf_t \xbf_t^T) - \eta \xbf_t ( \ \Hbf^{-1}(\xbf_t) \Vec{\nabla} g(\xbf_t) \ )^T  - \eta \Hbf^{-1}(\xbf_t) \Vec{\nabla} g(\xbf_t) \xbf_t^T +  \eta^2 (\Hbf^{-1}(\xbf_t) \Vec{\nabla} g(\xbf_t)  )  \cdot (\Hbf^{-1}(\xbf_t) \Vec{\nabla} g(\xbf_t)  )^T   ) $$
which is exactly the block encoding of $\xbf_{t+1} \xbf_{t+1}^T/(4p\Gamma^2) $ where we remind that $\Gamma^2 $ is $\sim$ one over minimum eigenvalue of $\Hbf(\xbf_t)$, or $||\Hbf(\xbf_t)^{-1}||$, which is known because we estimated it at first. This factor can be removed by amplification, resulting in further $\mathcal{O}(\Gamma^2)$ complexity. Therefore, the total running time of a single step of our quantum Newton's method is
$$ \mathcal{O}(  \mathcal{O}\Big(\gamma^{2p-2} 2s p^2 \ \big( p^2\log(n) + p\log^{2.5}(\frac{1}{\epsilon}) \big) \  \kappa \  polylog(\frac{\kappa}{\epsilon}) \Big)  ) $$

The improvement compared to~\cite{rebentrost2019quantum} turns out to be substantial as the number of copies requirement for $\xbf \xbf^T$ is significantly reduced. At a certain step of Newton's method, the work in~\cite{rebentrost2019quantum} requires $\mathcal{O}(p^5/\epsilon^3)$ copies of the temporal solution to perform an update of the solution, with a total time complexity $\mathcal{O}(p^8 \log(n) \kappa /\epsilon^4  )$. Meanwhile, the running time of our work depends polylogarithmically in terms of error tolerance, and that the number of ``copies'' of (block encoding of) $\xbf \xbf^T$ is $p$, which is a major advantage compared to~\cite{rebentrost2019quantum}. In particular, the dependence on $p$ reduces by a power-of-4. The most important thing is that our framework generalizes to polynomials of all kinds, not limited to homogeneous, even-degree ones. We finally emphasize that according to the analysis given in~\cite{rebentrost2019quantum}, the factor $\gamma$ from the above running time can be chosen to be arbitrarily small by choosing the parameter $\eta$ properly, e.g., choosing $\eta \leq 1/(2p ||\Hbf||^{-1})$ guarantees that $\gamma \leq 4$ (see Section 4, Result 4 of~\cite{rebentrost2014quantum} for detailed derivation). \\

An important aspect of our improved Newton's method is that the inversion of Hessian $\Hbf(\xbf)$ is more accurate. We recall that the method~\cite{childs2017quantum} requires a reasonable lower bound on the eigenvalues of $\Hbf(\xbf)$. However, in this case, the Hessian depends on $\xbf$, and it changes per each iteration. Therefore, in order to apply the method in~\cite{childs2017quantum}, we need to estimate the smallest eigenvalue in magnitude, which can be done efficiently, e.g., logarithmically with respect to the size of the matrix, using the method in~\cite{nghiem2023improved}. We note that this extra step has a running time smaller than the inversion itself, which implies that, asymptotically, it does not increase the overall complexity. In~\cite{rebentrost2019quantum}, the authors assumed the inversion is executed on some well-conditioned subspace of $\Hbf$, with the cutoff threshold chosen to be some constant $\Lambda_{H^{-1}}$. Where and how to obtain a reasonable value for $\Lambda_{H^{-1}}$ is unclear to us; as we mentioned, the spectrum of Hessian $\Hbf$ is not a constant over different iterations. Therefore, we believe that the extra step of estimating such a threshold is an important improvement upon~\cite{rebentrost2019quantum}.  

As another application of our convexity testing framework, in particular, one can see that our framework can be applied to enhance the performance of (improved) quantum Newton's method as well as gradient descent method. The reason is straightforward to see, as the gradient descent and quantum Newton's method are aimed to optimize a function by shifting toward the minima from some initialization. Therefore, by scanning the landscape, one can see if the given function is convex within such region. If the function is convex, then a minima is guaranteed, and hence, the algorithm is considered successful.

\section{Conclusion}
In this work, we have investigated the potential of quantum computers in the context of functional analysis, specifically testing the convexity of a given objective function. The problem of testing convexity is converted to the problem of determining the non-negativity of a matrix, e.g., all eigenvalues are non-negative. By employing a useful algebraic property of homogeneous polynomials of even degrees, we wrote down an explicit form of the so-called Hessian and built upon it to generalize the Hessian to arbitrary polynomials. We then combine it with a powerful quantum singular value transformation framework, plus a quantum power method to construct a quantum algorithm that allows us to test the positive-semidefiniteness of such Hessian matrix at multiple points from a given domain. The procedure has running time polylogarithmically respective to the dimension $n$ and linear with the number of sample points $\Nsr$, which is superpolynomially faster than the best-known classical approach relative to the number of variables, meanwhile keeping the same complexity on the number of sample points. We also point out two examples where we envision a potential application of convexity testing, including studying the geometric structure of manifold, testing training landscape of variational quantum algorithms as well as optimization landscape for gradient descent/Newton's method for optimization. In particular, as a striking corollary of our result, we provide a major improvement upon the work of~\cite{rebentrost2019quantum} in multiple aspects: running time (with respect to error tolerance), generality (arbitrary polynomial type) and the subtle detail regarding the inversion of the Hessian. As a whole, our work has added one more interesting example to the field of quantum computation, revealing that the area of functional analysis, and many more areas, is a rich avenue deserving of further research from computational aspects. An interesting open direction that we believe is worth looking at is that the model we work on in this case is somewhat explicit, e.g., the function is a polynomial of computable coefficients given via an oracle. In some traditional problems, the input function is typically given as a blackbox, such as Grover's search problem~\cite{grover1996fast}, and we wish to reveal hidden properties. Suppose, instead, we are given a blackbox function that computes some analytical function; then, how do we extract the hidden properties, such as the convexity of the function in some domain? We note that there are two prior relevant works such as~\cite{jordan2005fast}, where the author considered the gradient estimation, and~\cite{childs2022quantum}, where the authors considered a sampling problem from a blackbox oracle that computes the value of given function within some domain. While there are seemingly overlaps, we do not see a direct solution to our case, thereby we leave the challenge for future investigation.

\section*{Acknowledgement}
We thank Hiroki Sukeno and Shuyu Zhang for carefully reading and detailed feedback on the manuscript. This work was supported by the U.S. Department of Energy, Office of Science, Advanced
Scientific Computing Research under Award Number DE-SC-0012704. 
We also acknowledge the support of a Seed Grant from
Stony Brook University’s Office of the Vice President for Research and the Center for Distributed Quantum Processing.

\bibliography{ref.bib}{}

\begin{thebibliography}{10}

\bibitem{deutsch1985quantum}
David Deutsch.
\newblock Quantum theory, the church--turing principle and the universal quantum computer.
\newblock {\em Proceedings of the Royal Society of London. A. Mathematical and Physical Sciences}, 400(1818):97--117, 1985.

\bibitem{deutsch1992rapid}
David Deutsch and Richard Jozsa.
\newblock Rapid solution of problems by quantum computation.
\newblock {\em Proceedings of the Royal Society of London. Series A: Mathematical and Physical Sciences}, 439(1907):553--558, 1992.

\bibitem{shor1994proceedings}
Peter~W Shor.
\newblock Proceedings of the 35th annual symposium on foundations of computer science.
\newblock {\em IEE Computer society press, Santa Fe, NM}, 1994.

\bibitem{regev2023efficient}
Oded Regev.
\newblock An efficient quantum factoring algorithm.
\newblock {\em arXiv preprint arXiv:2308.06572}, 2023.

\bibitem{grover1996fast}
Lov~K Grover.
\newblock A fast quantum mechanical algorithm for database search.
\newblock In {\em Proceedings of the twenty-eighth annual ACM symposium on Theory of computing}, pages 212--219, 1996.

\bibitem{berry2007efficient}
Dominic~W Berry, Graeme Ahokas, Richard Cleve, and Barry~C Sanders.
\newblock Efficient quantum algorithms for simulating sparse hamiltonians.
\newblock {\em Communications in Mathematical Physics}, 270(2):359--371, 2007.

\bibitem{berry2012black}
Dominic~W Berry and Andrew~M Childs.
\newblock Black-box hamiltonian simulation and unitary implementation.
\newblock {\em Quantum Information and Computation}, 12:29--62, 2009.

\bibitem{berry2014high}
Dominic~W Berry.
\newblock High-order quantum algorithm for solving linear differential equations.
\newblock {\em Journal of Physics A: Mathematical and Theoretical}, 47(10):105301, 2014.

\bibitem{berry2015hamiltonian}
Dominic~W Berry, Andrew~M Childs, and Robin Kothari.
\newblock Hamiltonian simulation with nearly optimal dependence on all parameters.
\newblock In {\em 2015 IEEE 56th annual symposium on foundations of computer science}, pages 792--809. IEEE, 2015.

\bibitem{low2017optimal}
Guang~Hao Low and Isaac~L Chuang.
\newblock Optimal hamiltonian simulation by quantum signal processing.
\newblock {\em Physical review letters}, 118(1):010501, 2017.

\bibitem{low2019hamiltonian}
Guang~Hao Low and Isaac~L Chuang.
\newblock Hamiltonian simulation by qubitization.
\newblock {\em Quantum}, 3:163, 2019.

\bibitem{harrow2009quantum}
Aram~W Harrow, Avinatan Hassidim, and Seth Lloyd.
\newblock Quantum algorithm for linear systems of equations.
\newblock {\em Physical review letters}, 103(15):150502, 2009.

\bibitem{childs2017quantum}
Andrew~M Childs, Robin Kothari, and Rolando~D Somma.
\newblock Quantum algorithm for systems of linear equations with exponentially improved dependence on precision.
\newblock {\em SIAM Journal on Computing}, 46(6):1920--1950, 2017.

\bibitem{lloyd2013quantum}
Seth Lloyd, Masoud Mohseni, and Patrick Rebentrost.
\newblock Quantum algorithms for supervised and unsupervised machine learning.
\newblock {\em arXiv preprint arXiv:1307.0411}, 2013.

\bibitem{mitarai2018quantum}
Kosuke Mitarai, Makoto Negoro, Masahiro Kitagawa, and Keisuke Fujii.
\newblock Quantum circuit learning.
\newblock {\em Physical Review A}, 98(3):032309, 2018.

\bibitem{lloyd2014quantum}
Seth Lloyd, Masoud Mohseni, and Patrick Rebentrost.
\newblock Quantum principal component analysis.
\newblock {\em Nature Physics}, 10(9):631--633, 2014.

\bibitem{lloyd2016quantum}
Seth Lloyd, Silvano Garnerone, and Paolo Zanardi.
\newblock Quantum algorithms for topological and geometric analysis of data.
\newblock {\em Nature communications}, 7(1):1--7, 2016.

\bibitem{huang2022quantum}
Hsin-Yuan Huang, Michael Broughton, Jordan Cotler, Sitan Chen, Jerry Li, Masoud Mohseni, Hartmut Neven, Ryan Babbush, Richard Kueng, John Preskill, et~al.
\newblock Quantum advantage in learning from experiments.
\newblock {\em Science}, 376(6598):1182--1186, 2022.

\bibitem{jordan2005fast}
Stephen~P Jordan.
\newblock Fast quantum algorithm for numerical gradient estimation.
\newblock {\em Physical review letters}, 95(5):050501, 2005.

\bibitem{rebentrost2019quantum}
Patrick Rebentrost, Maria Schuld, Leonard Wossnig, Francesco Petruccione, and Seth Lloyd.
\newblock Quantum gradient descent and newton’s method for constrained polynomial optimization.
\newblock {\em New Journal of Physics}, 21(7):073023, 2019.

\bibitem{rebentrost2014quantum}
Patrick Rebentrost, Masoud Mohseni, and Seth Lloyd.
\newblock Quantum support vector machine for big data classification.
\newblock {\em Physical review letters}, 113(13):130503, 2014.

\bibitem{rebentrost2018quantum}
Patrick Rebentrost, Thomas~R Bromley, Christian Weedbrook, and Seth Lloyd.
\newblock Quantum hopfield neural network.
\newblock {\em Physical Review A}, 98(4):042308, 2018.

\bibitem{schuld2018supervised}
Maria Schuld and Francesco Petruccione.
\newblock {\em Supervised learning with quantum computers}, volume~17.
\newblock Springer, 2018.

\bibitem{prakash2014quantum}
Anupam Prakash.
\newblock {\em Quantum algorithms for linear algebra and machine learning}.
\newblock University of California, Berkeley, 2014.

\bibitem{gilyen2019quantum}
Andr{\'a}s Gily{\'e}n, Yuan Su, Guang~Hao Low, and Nathan Wiebe.
\newblock Quantum singular value transformation and beyond: exponential improvements for quantum matrix arithmetics.
\newblock In {\em Proceedings of the 51st Annual ACM SIGACT Symposium on Theory of Computing}, pages 193--204, 2019.

\bibitem{gilyen2019quantum1}
Andr{\'a}s Gily{\'e}n.
\newblock {\em Quantum singular value transformation \& its algorithmic applications}.
\newblock PhD thesis, University of Amsterdam, 2019.

\bibitem{nghiem2023improved}
Nhat~A Nghiem and Tzu-Chieh Wei.
\newblock Improved quantum algorithms for eigenvalues finding and gradient descent.
\newblock {\em arXiv preprint arXiv:2312.14786}, 2023.

\bibitem{ccevik2011spectrum}
E~Otkun {\c{C}}evik and Zameddin~I Ismailov.
\newblock Spectrum of the direct sum of operators.
\newblock {\em arXiv preprint arXiv:1105.4223}, 2011.

\bibitem{cerezo2021variational}
Marco Cerezo, Andrew Arrasmith, Ryan Babbush, Simon~C Benjamin, Suguru Endo, Keisuke Fujii, Jarrod~R McClean, Kosuke Mitarai, Xiao Yuan, Lukasz Cincio, et~al.
\newblock Variational quantum algorithms.
\newblock {\em Nature Reviews Physics}, 3(9):625--644, 2021.

\bibitem{mcclean2016theory}
Jarrod~R McClean, Jonathan Romero, Ryan Babbush, and Al{\'a}n Aspuru-Guzik.
\newblock The theory of variational hybrid quantum-classical algorithms.
\newblock {\em New Journal of Physics}, 18(2):023023, 2016.

\bibitem{havlivcek2019supervised}
Vojt{\v{e}}ch Havl{\'\i}{\v{c}}ek, Antonio~D C{\'o}rcoles, Kristan Temme, Aram~W Harrow, Abhinav Kandala, Jerry~M Chow, and Jay~M Gambetta.
\newblock Supervised learning with quantum-enhanced feature spaces.
\newblock {\em Nature}, 567(7747):209--212, 2019.

\bibitem{peruzzo2014variational}
Alberto Peruzzo, Jarrod McClean, Peter Shadbolt, Man-Hong Yung, Xiao-Qi Zhou, Peter~J Love, Al{\'a}n Aspuru-Guzik, and Jeremy~L O’brien.
\newblock A variational eigenvalue solver on a photonic quantum processor.
\newblock {\em Nature communications}, 5(1):4213, 2014.

\bibitem{farhi2014quantum}
Edward Farhi, Jeffrey Goldstone, and Sam Gutmann.
\newblock A quantum approximate optimization algorithm.
\newblock {\em arXiv preprint arXiv:1411.4028}, 2014.

\bibitem{wang2021noise}
Samson Wang, Enrico Fontana, Marco Cerezo, Kunal Sharma, Akira Sone, Lukasz Cincio, and Patrick~J Coles.
\newblock Noise-induced barren plateaus in variational quantum algorithms.
\newblock {\em Nature communications}, 12(1):6961, 2021.

\bibitem{cerezo2021cost}
Marco Cerezo, Akira Sone, Tyler Volkoff, Lukasz Cincio, and Patrick~J Coles.
\newblock Cost function dependent barren plateaus in shallow parametrized quantum circuits.
\newblock {\em Nature communications}, 12(1):1791, 2021.

\bibitem{bittel2021training}
Lennart Bittel and Martin Kliesch.
\newblock Training variational quantum algorithms is np-hard.
\newblock {\em Physical review letters}, 127(12):120502, 2021.

\bibitem{lloyd2020quantum}
Seth Lloyd, Maria Schuld, Aroosa Ijaz, Josh Izaac, and Nathan Killoran.
\newblock Quantum embeddings for machine learning.
\newblock {\em arXiv preprint arXiv:2001.03622}, 2020.

\bibitem{childs2022quantum}
Andrew~M Childs, Tongyang Li, Jin-Peng Liu, Chunhao Wang, and Ruizhe Zhang.
\newblock Quantum algorithms for sampling log-concave distributions and estimating normalizing constants.
\newblock {\em Advances in Neural Information Processing Systems}, 35:23205--23217, 2022.

\bibitem{camps2020approximate}
Daan Camps and Roel Van~Beeumen.
\newblock Approximate quantum circuit synthesis using block encodings.
\newblock {\em Physical Review A}, 102(5):052411, 2020.

\bibitem{childs2017lecture}
Andrew~M Childs.
\newblock Lecture notes on quantum algorithms.
\newblock {\em Lecture notes at University of Maryland}, 2017.

\end{thebibliography}
\bibliographystyle{unsrt}

\clearpage
\newpage
\onecolumngrid
\appendix
\section{Preliminaries}
\label{sec: prelim}
Here, we summarize the main recipes of our work. We keep their statements brief but precise for simplicity, with their proofs/ constructions referred to in their original works.

\begin{definition}[Block Encoding Unitary]~\cite{low2017optimal, low2019hamiltonian, gilyen2019quantum}
\label{def: blockencode} 
Let $A$ be some Hermitian matrix of size $N \times N$ whose matrix norm $|A| < 1$. Let a unitary $U$ have the following form:
\begin{align*}
    U = \begin{pmatrix}
       A & \cdot \\
       \cdot & \cdot \\
    \end{pmatrix}.
\end{align*}
Then $U$ is said to be an exact block encoding of matrix $A$. Equivalently, we can write:
\begin{align*}
    U = \ket{ \bf{0}}\bra{ \bf{0}} \otimes A + \cdots,
\end{align*}
where $\ket{\bf 0}$ refers to the ancilla system required for the block encoding purpose. In the case where the $U$ has the form 
$$ U  =  \ket{ \bf{0}}\bra{ \bf{0}} \otimes \Tilde{A} + \cdots, $$
where $|| \Tilde{A} - A || \leq \epsilon$ (with $||.||$ being the matrix norm), then $U$ is said to be an $\epsilon$-approximated block encoding of $A$.
\end{definition}

The above definition has multiple simple corollaries. First, an arbitrary unitary $U$ block encodes itself. Suppose $A$ is block encoded by some matrix $U$, then $A$ can be block encoded in a larger matrix by simply adding ancillas (which have dimension $m$). Note that $\Ibb_m \otimes U$ contains $A$ in the top-left corner, which is a block encoding of $A$ again by definition. Further, it is almost trivial to block encode the identity matrix of any dimension. For instance, we consider $\sigma_z \otimes \Ibb_m$ (for any $m$), which contains $\Ibb_m$ in the top-left corner.

\begin{lemma}[\cite{gilyen2019quantum}]
\label{lemma: improveddme}
Let $\rho = \Tr_A \ket{\Phi}\bra{\Phi}$, where $\rho \in \mathbb{H}_B$, $\ket{\Phi} \in  \mathbb{H}_A \otimes \mathbb{H}_B$. Given unitary $U$ that generates $\ket{\Phi}$ from $\ket{\bf 0}_A \otimes \ket{\bf 0}_B$, then there exists an efficient procedure that constructs an exact unitary block encoding of $\rho$.
\end{lemma}

The proof of the above lemma is given in \cite{gilyen2019quantum} (see their Lemma 45). \\

\begin{lemma}[Block Encoding of Product of Two Matrices]
\label{lemma: product}
    Given the unitary block encoding of two matrices $A_1$ and $A_2$, an efficient procedure exists that constructs a unitary block encoding of $A_1 A_2$.
\end{lemma}

The proof of the above lemma is also given in~\cite{gilyen2019quantum}.  \\

\begin{lemma}[\cite{camps2020approximate}]
\label{lemma: tensorproduct}
    Given the unitary block encoding $\{U_i\}_{i=1}^m$ of multiple operators $\{M_i\}_{i=1}^m$ (assumed to be exact encoding), then, there is a procedure that produces the unitary block encoding operator of $\bigotimes_{i=1}^m M_i$, which requires a single use of each $\{U_i\}_{i=1}^m$ and $\mathcal{O}(1)$ SWAP gates. 
\end{lemma}
The above lemma is a result in~\cite{camps2020approximate}. 
\begin{lemma}
\label{lemma: As}
    Given the oracle access to $s$-sparse matrix $A$ of dimension $n\times n$, then an $\epsilon$-approximated unitary block encoding of $A/s$ can be prepared with gate/time complexity $\mathcal{O}(\log n + \log^{2.5}(\frac{1}{\epsilon}))$.
\end{lemma}
This is also presented in~\cite{gilyen2019quantum}.  One can also find similar construction in Ref.~\cite{childs2017lecture}. 
\begin{lemma}
    Given unitary block encoding of multiple operators $\{M_i\}_{i=1}^m$. Then, there is a procedure that produces a unitary block encoding operator of $\sum_{i=1}^m \pm M_i/m $ in complexity $\mathcal{O}(m)$.
    \label{lemma: sumencoding}
\end{lemma}

\begin{lemma}[Scaling Block encoding]
\label{lemma: scale}
    Given a block encoding of some matrix $A$ (as in~\ref{def: blockencode}), then the block encoding of $A/p$, where $p > 1$, can be prepared with an extra $\mathcal{O}(1)$ cost.
\end{lemma}

\section{Proof of Lemma~\ref{lemma: alphadiagonal}}
\label{sec: alphadiagonal}
Remind that we are trying to prove the following
\begin{lemma}
   Given the block encoding of $\bigoplus_{i=1}^\Nsr cc^T = \Ibb_\Nsr cc^T$ and $\bigoplus_{i=1}^\Nsr \xbf_i\xbf_i^T$, then it is able to constructs the block encoding of the diagonal matrix $\mathscr{B}$ with entries $\mathscr{B}_{ij} = \beta_i^2  \delta_{ij}$ where $\beta_i = \xbf_i^T c$
\end{lemma}
From the block encoding of $\bigoplus_{i=1}^\Nsr cc^T = \Ibb_\Nsr cc^T$ as in the above lemma, plus the block encoding of $\bigoplus_{i=1}^{\Nsr} \xbf_i \xbf_i^T$, we can use lemma \ref{lemma: product} to construct the block encoding of their products, e.g, we would obtain the block encoding of $\bigoplus_{i=1}^\Nsr \beta_i c \xbf_i^T$, denoted as $U_\beta$. Given the definition of block encoding (\ref{def: blockencode}), we have that:
\begin{align}
    U_\beta \ket{\bf 0}_u \ket{\Phi} = \ket{\bf 0}_u (\bigoplus_{i=1}^\Nsr \beta_i c \xbf_i^T) \ket{\Phi} + \ket{Garbage}
\end{align}
where $\ket{Garbage}$ satisfies: $\ket{\bf 0}\bra{\bf 0} \otimes \Ibb \cdot \ket{Garbage} = 0$. Given that $\bigoplus_{i=1}^\Nsr \beta_i c \xbf_i^T = \sum_{i=1}^\Nsr \ket{i}\bra{i} \otimes \beta_i c \xbf_i^T$ and if we choose $\ket{\Phi} =  \ket{j} \otimes \ket{c}$, we have that 
$$ \bigoplus_{i=1}^\Nsr \beta_i c \xbf_i^T \cdot \ket{\Phi} = \delta_{ij} \beta_i \beta_j  c $$
Recall further that we also have a unitary $U_C$ such that $U_C\ket{\bf 0} = \ket{c} \equiv c$. Therefore, the state $\ket{\Phi}$ could be created from $\ket{\bf 0}\otimes \ket{j}$ by simply applying $U_C$ to the first register, e.g, to obtain $\ket{c} \otimes \ket{j}$ and use SWAP gate to swap them, e.g, we obtain $\ket{j} \otimes \ket{c}$. Denote such process as $U_s$.

To match the same dimension as $U_\beta$, we simply add extra register (with corresponding dimension). We have that:
\begin{align}
 \bra{\bf 0}_u \bra{\bf 0}\bra{i}  (\Ibb_u \otimes U_C^\dagger) \  U_\beta \ (\Ibb_u \otimes U_s) \ket{\bf 0}_u \ket{\bf 0}\ket{j} = \delta_{ij} \beta_i \beta_j
\end{align}
which basically is a block encoding of a diagonal matrix that contains $\beta_i^2$. Using the result from \cite{gilyen2019quantum} to remove the power factor $2$, we are left with block encoding of $\delta_{ij} \beta_i \beta_j$.

\end{document}